\documentclass[twocolumn,preprintnumbers,superscriptaddress,nofootinbib,aps,prd,floatfix]{revtex4}

\usepackage{amsmath,amssymb}
\usepackage{dsfont}
\usepackage{graphicx} 
\usepackage{epstopdf} 
\usepackage{slashed}
\usepackage{subfigure}
\usepackage{color}
\usepackage{multirow}
\usepackage{hyperref}

\usepackage{footmisc}

\usepackage{array}
\newcolumntype{L}[1]{>{\raggedright\let\newline\\\arraybackslash\hspace{0pt}}m{#1}}
\newcolumntype{C}[1]{>{\centering\let\newline\\\arraybackslash\hspace{0pt}}m{#1}}
\newcolumntype{R}[1]{>{\raggedleft\let\newline\\\arraybackslash\hspace{0pt}}m{#1}}

\def\sst{\scriptscriptstyle}

\newcommand{\be}{\begin{eqnarray*}}
\newcommand{\ee}{\end{eqnarray*}}

\newcommand{\bee}{\begin{eqnarray}}
\newcommand{\eee}{\end{eqnarray}}
\newcommand{\beeq}{\begin{equation}}
\newcommand{\eeq}{\end{equation}}

\renewcommand{\vec}{\bf}

\usepackage{xspace}

\usepackage{color}
\usepackage{xcolor}

\begin{document}

\title{Hearing the signals of dark sectors with gravitational wave detectors}

\begin{abstract}
\noindent Motivated by aLIGO's recent discovery of gravitational waves we discuss signatures of new physics that could be seen at ground and space-based interferometers. We show that a first order phase transition in a dark sector would lead to a detectable
gravitational wave signal at future experiments, if the phase transition has occurred at temperatures few orders of magnitude higher than the electroweak scale. The source of gravitational waves in this case is associated with the dynamics of expanding and colliding bubbles in the early universe.
At the same time we point out that topological defects, such as dark sector domain walls, may generate a detectable signal already at aLIGO. Both -- bubble and domain wall -- scenarios are sourced by semi-classical configurations of a dark new physics sector.
In the first case the gravitational wave signal originates from bubble wall collisions and subsequent turbulence in hot plasma in the early universe,
while the second case corresponds to domain walls passing through the interferometer at present and is not related to gravitational waves. 
We find that aLIGO at its current sensitivity can detect smoking-gun signatures from domain wall interactions, while future proposed experiments including the fifth phase of aLIGO at design sensitivity can probe dark sector phase transitions.
\end{abstract}
\author{Joerg Jaeckel} \email{jjaeckel@thphys.uni-heidelberg.de}
\affiliation{Institut f\"ur theoretische Physik, Universit\"at Heidelberg, Philosophenweg 16, 69120 Heidelberg, Germany\\[0.1cm]}
\author{Valentin V. Khoze} \email{valya.khoze@durham.ac.uk}
\affiliation{Institute for Particle Physics Phenomenology, Department 
  of Physics,\\Durham University, DH1 3LE, United Kingdom\\[0.1cm]}
\author{Michael Spannowsky$^{2}$} \email{michael.spannowsky@durham.ac.uk}

\preprint{IPPP/16/12,  DCPT/16/24}
\maketitle

\section{Introduction}
\label{sec:intro}
The sublime discovery of gravitational waves at advanced LIGO (aLIGO) \cite{aLIGO} is yet another striking confirmation of Einstein's theory of gravity. Due to the weakness of gravitational interactions and the fact that gravity couples to all particles that carry energy and momentum, gravitational waves (GW) are at the same time witness to and remnant of some of the most violent phenomena in our Universe, e.g. Neutron-star inspirals, Black Hole inspirals, Pulsars or phase transitions. They herald intense dynamics, potentially from a distant past.

In recent years, a strong effort was made to discover gravitational waves using ground-based experiments. After  somewhat uneventful runs of, for example, LIGO \cite{Abramovici:1992ah}, Virgo \cite{Giazotto:1988gw}, or the European Pulsar Timing Array (EPTA) \cite{Ferdman:2010xq}, in 2015 aLIGO \cite{Harry:2010zz} started operations with increased sensitivity in gravitational wave frequencies of $10^0$-$10^3$ Hz and a reach well into the characteristic strain of supernovae, pulsars and binary inspirals. 

While aLIGO was primarily designed to detect gravitational waves from a multitude of astrophysical sources, it retains a remarkable sensitivity to new physics effects. Adding gravitational wave detection experiments as an additional arrow to the quiver of experiments to search for new physics interactions will help to probe very weakly coupled sectors of new physics. 

With obvious short-comings in our understanding of fundamental principles of nature dangling, e.g. the lack of a dark matter candidate or the observed matter/anti-matter asymmetry, and in absence of evidence for new physics at collider experiments, so-called dark sectors become increasingly attractive as add-on to the Standard Model. If uncharged under the Standard Model gauge group, dark sectors could even have a rich particle spectrum without leaving an observable imprint in measurements at particle colliders. Hence, this could leave us in the strenuous situation where we might have to rely exclusively on very feeble possibly only gravitational interactions to infer their existence. 

For dark sectors to address the matter/anti-matter asymmetry via electroweak baryogenesis, usually a strong first-order phase transition is required\footnote{For an interesting recent mechanism to do baryogenesis with dark sector phase transitions see~\cite{Katz:2016adq}.}. It is well known that a first-order phase transition is accompanied by three mechanisms that can give rise to gravitational waves 
in the early universe 
\cite{Kosowsky:1991ua,Kamionkowski:1993fg,Grojean:2006bp,Huber:2008hg,Caprini:2009yp,Binetruy:2012ze,Hindmarsh:2015qta,Caprini:2015zlo}: collisions of expanding vacuum bubbles, sounds waves, and magnetohydrodynamic turbulence of bubbles in the hot plasma.
However, for previously studied models, e.g. (N)MSSM \cite{Huber:2015znp},
strongly coupled dark sectors \cite{Schwaller:2015tja}, or the electroweak phase transition with the Higgs potential modified by a sextic term \cite{Huang:2016odd},
the resulting GW frequencies after red-shifting are expected to have frequencies of some two or more orders of magnitude below the reach of aLIGO. 
 On the other hand, if electroweak symmetry breaking is triggered in the dark sector at temperatures significantly above the electroweak scale, e.g. by radiatively generating a vev using the Coleman-Weinberg mechanism, GW with frequencies are within the aLIGO reach, i.e. 1-100 Hz. 
However, we will explain that the overall amplitude of the signal is too small for aLIGO at present sensitivity, but it can be probed by the next generation of interferometers.\footnote{These future experiments also include the advanced LIGO/VIRGO detectors operating in years 2020+ at the projected final sensitivity 
\cite{TheLIGOScientific:2016wyq}
as was also pointed out very recently in \cite{Dev:2016feu}.}  

At the same time, already now, aLIGO can probe beyond the standard model physics. We will investigate the consequences of topological defects, such as a domain wall passing through the interferometer. We will model this by introducing a non-vanishing effective photon mass localised on the domain wall, while vanishing elsewhere.\footnote{This is not a gravitational effect, but effectively it looks like local ripples affecting the propagation of photons.} 
The signatures of passing domain walls can be well separated from black-hole mergers and motivates an extension of ongoing search strategies. 

In Sec.~\ref{sec:PT} we discuss the implementation of first order phase transitions in dark sectors with radiative symmetry breaking. Sec.~\ref{sec:domain} is dedicated to the modelling and phenomenology of the domain wall interacting with aLIGO. We offer a summary in Sec.~\ref{sec:conclusion}.

\section{First-order phase transition in a dark sector at high scales}
\label{sec:PT}

\subsection{Dark sector model at zero temperature}

Let us consider a very simple minimal model of the hidden (or dark sector) consisting of a complex scalar $\Phi$ 
which is a SM singlet, i.e. it does not couple to any of the Standard Model gauge groups but is
charged under the gauge group of the dark sector -- in the simplest case a U(1) gauge group.
The SM Higgs doublet $H$
is coupled via the Higgs-portal interactions to the complex scalar 
\begin{equation}
\Phi \,=\, \frac{1}{\sqrt{2}}(\phi + i \phi_2)\,.
\end{equation}
In unitary gauge one is left with two real scalars,
\begin{equation}
H=\frac{1}{\sqrt{2}}(0,h)\, , \quad 
\Phi=\frac{1}{\sqrt{2}}\phi\,,
\end{equation}
and the tree-level scalar potential reads
\begin{equation}
V_0(h,\phi)=\frac{\lambda_{\phi}}{4}\phi^4+\frac{\lambda_H}{4}h^4-\frac{\lambda_{\rm P}}{4} h^2 \phi^2\,.
\label{V0hphi}
\end{equation}
Note that we have assumed that the theory is scale-invariant at the classical level \cite{Coleman:1973jx}, and as the result,
none of the mass scales are present in the theory, they can only be generated quantum mechanically i.e. via radiative corrections.
(Of course, one can also consider more general examples of hidden sectors, which are not classically scale invariant and still have first order phase transitions.)
\begin{figure}[t]
\includegraphics[width=0.43\textwidth]{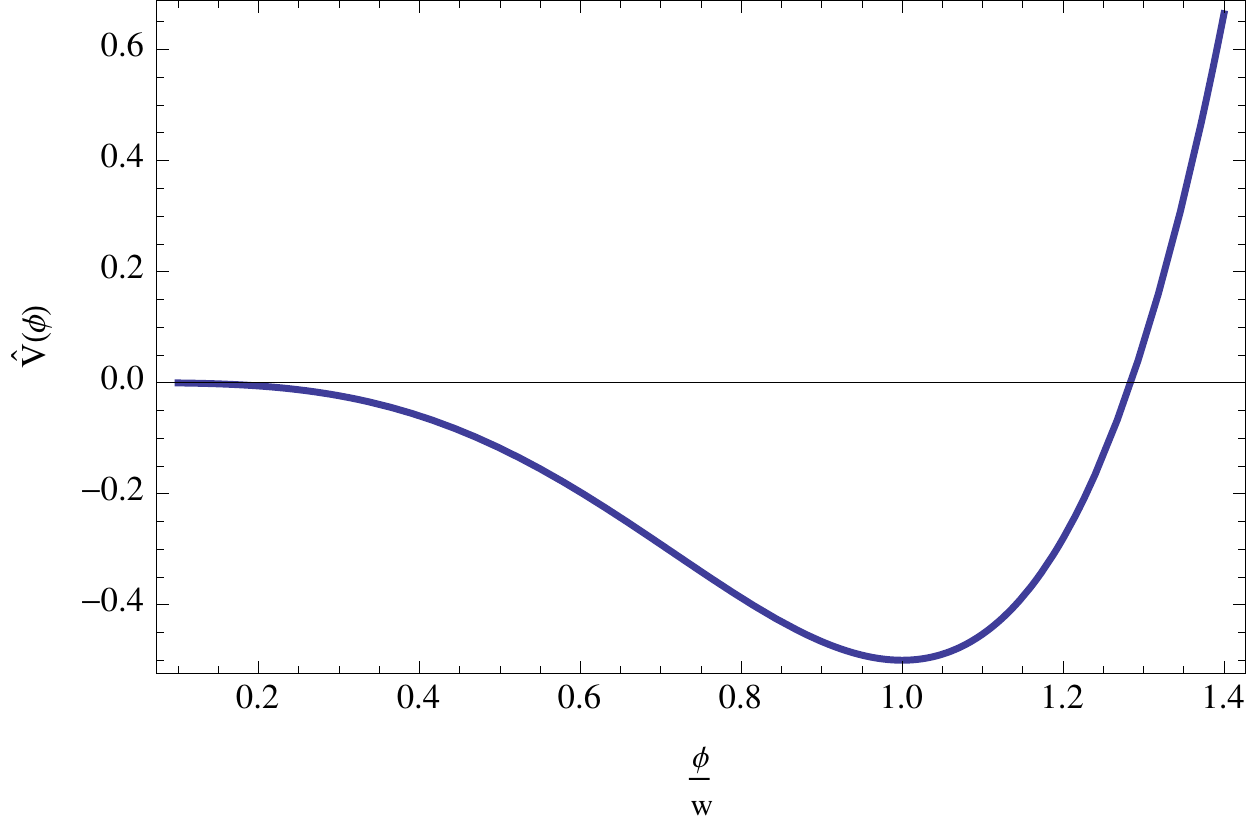}
  \caption{\small The zero-temperature effective potential $V$ of the CW theory Eq. \eqref{V1RR} in the units of
  $\frac{3}{64 \pi ^2} \, g_{\sst \mathrm{D}}^4$.}
  \label{fig:VT0}
\end{figure}

In the minimal Standard Model classical scale invariance is broken by the Higgs mass parameter
$\mu^{2}_{\sst \mathrm{SM}}$ . 
Scale invariance is easily restored by reinterpreting this scale
in terms of the vev of a $\phi$, coupled to the SM via the
Higgs portal interaction, $-\,(\lambda_{\rm P}/4)h^2\phi^{2}$ in \eqref{V0hphi}.
Now, as soon as an appropriate non-vanishing value for $\langle \phi\rangle\ll M_{\sst \mathrm{UV}}$ is generated (as we will see momentarily), we get  
$\mu^2_{\sst \mathrm{SM}} = \lambda_{\rm P}\langle|\phi|\rangle^2$ 
which triggers electroweak symmetry breaking. (For more detail on this see a recent discussion in \cite{Englert:2013gz,Khoze:2014xha} and references therein.)

From now on we will concentrate on the dark sector alone and neglect the 
back reaction of the SM; these corrections can be straightforwardly included, but will not be essential to our discussion.
The zero-temperature 1-loop effective potential for $\phi$ reads \cite{Coleman:1973jx},
\begin{equation}
V (\phi;\mu)\,=\, 
\frac{\lambda_\phi(\mu)}{4}\phi^4+ \frac{n g_{\sst \mathrm{D}}(\mu)^4}{64 \pi ^2}  \phi^4
\left(\log \left(\frac{\phi^2}{\mu^2}\right)-\frac{25}{6}\right) \,,
\label{V1R}
\end{equation}
where $\mu$ is the RG scale, $g_{\sst \mathrm{D}}$ is the U(1) dark sector gauge coupling, and the
second term on the {\it r.h.s.} are the 1-loop contributions arising from the hidden U(1) gauge bosons $Z'$.
In this case the factor of $n$ appearing on the {\it r.h.s} of \eqref{V1R} is $n=3$.
The vacuum of the effective potential above occurs at $\langle \phi \rangle \neq 0.$ 
Minimising the potential \eqref{V1R} with respect to $\phi$ at $\mu=\langle \phi \rangle$ gives the 
characteristic Coleman-Weinberg-type $\lambda_\phi \propto g_{\sst \mathrm{CW}}^4$ relation between 
the scalar and the gauge couplings,
\begin{equation}
\lambda_\phi \,=\, \frac{11}{16\pi^2} \,g_{\sst \mathrm{D}}^4 \qquad {\rm at} \quad \mu=\langle \phi\rangle \equiv w\,.
\label{eq:cwmsbar}
\end{equation}
From now on we will refer to the non-vanishing vev of $\phi$ in the zero-temperature theory as $w$.
With this matching condition at $\mu=w$ the zero-temperature effective potential \eqref{V1R}
for the U(1) CW theory takes the form,
\begin{equation}
V (\phi)\,=\, 
\frac{n}{64 \pi ^2} \, g_{\sst \mathrm{D}}^4\, \phi^4
\left(-\frac{1}{2}+\log \left(\frac{\phi^2}{w^2}\right)\right).
\label{V1RR}
\end{equation}
It is plotted in Fig.~\ref{fig:VT0} which shows the existence of a single vacuum at $\phi=w$ generated via radiative 
corrections.
The physical mass of the CW scalar is found by expanding \eqref{V1RR} around $\phi \to w +\phi$,
\begin{equation}
m_\phi^2\,=\,
\frac{ng_{\sst \mathrm{D}}^4}{8\pi^2}w^2\,,
\label{eq:mphiZ}
\end{equation}
and the mass of the $Z'$ vector boson is
$M_{Z'} = \frac{1}{2}g_{\sst \mathrm{D}} w \gg m_\phi.$

The above formulae are easily generalised also to non-Abelian CW gauge groups. 
For example in a classically scale-invariant SU(2) gauge theory with the scalar field in the adjoint representation considered e.g. in \cite{Khoze:2014woa}
one just sets $n=6$
and hence
\begin{equation}
V (\phi)\,=\, 
\frac{6}{64 \pi ^2} \, g_{\sst \mathrm{D}}^4\, \phi^4
\left(-\frac{1}{2}+\log \left(\frac{\phi^2}{w^2}\right)\right) \,.
\label{V2RR}
\end{equation}
The only difference between \eqref{V1RR} and \eqref{V2RR} is that in the SU(2) case there are two $W'$ bosons contributing to the
loops, hence the total of 6 degrees of freedom compared to 3 on the {\it r.h.s.} of \eqref{V1RR}.

In the rest of this section we will concentrate on the SU(2) with the adjoint scalar case in hand, i.e. $n=6$. 
One can also easily switch to the U(1) theory conventions, and other examples of CW hidden sectors, such as 
the SU(2) with the scalar in the fundamental representation, and the U(1)$_{B-L}$ classically scale-invariant extensions 
of the Standard Model were considered in \cite{Khoze:2014xha}.

\subsection{Thermal effects}
\label{sec:thermal}

\begin{figure}[t]
\includegraphics[width=0.43\textwidth]{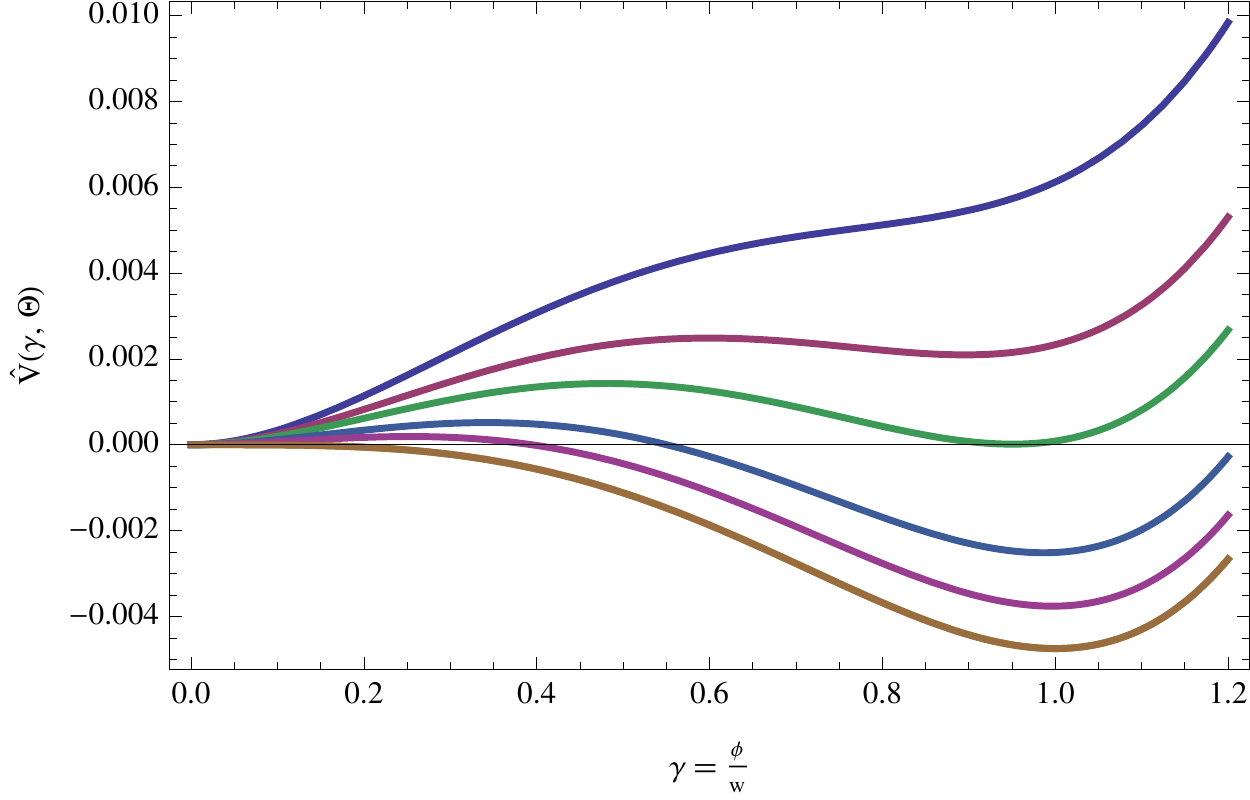}
  \caption{\small Thermal effective potential $\hat{V}(\gamma,\Theta)$ of the dark sector in Eq. \eqref{eq:VT1} as a function 
  of $\gamma=\phi/w$ plotted for different 
  temperatures $\Theta=$ 0.40, 0.35, 0.31, 0.25, 0.20 and 0 (from top to bottom).
  We have shifted $\hat{V}(\gamma,\Theta)$ by a constant so that the effective potential 
  at the origin is zero for all values of $\Theta$.}
  \label{fig:VT1}
\end{figure}
\begin{figure}[t]
\includegraphics[width=0.43\textwidth]{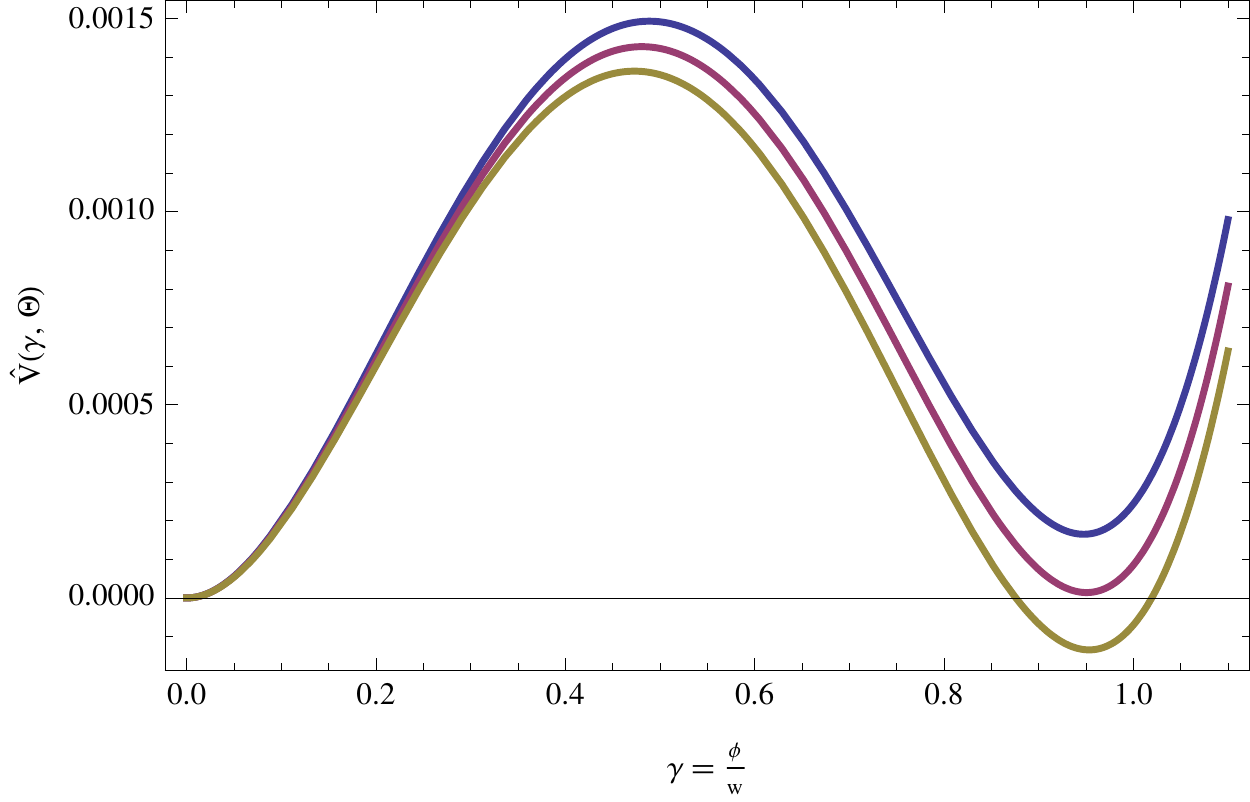}
  \caption{\small Thermal effective potential $\hat{V}(\gamma,\Theta)$ as in Fig~\ref{fig:VT1} now zooming
  at the values around the critical 
  temperature,
   $\Theta=$ 0.315, 0.312, 0.309 (from top to bottom).
 }
  \label{fig:VT2}
\end{figure}
The effective potential at finite
temperature along the $\phi$ direction is given by the zero-temperature effective potential \eqref{V2RR}
plus the purely thermal correction $\Delta V_{T}$ which vanishes at $T=0$,
\begin{equation}
V_{T}(\phi)=\, V(\phi)+ \Delta V_{T}(\phi)\,.
\label{Vth0}
\end{equation}
The second term is computed
at one-loop in perturbation theory and is given by the well-known expression\cite{Dolan:1973qd}:
\begin{equation}
\frac{T^{4}}{2\pi^{2}}\sum_{i}\pm n_{i}\int_{0}^{\infty}\mbox{d}q\, 
q^{2}\log\left(1\mp\exp(-\sqrt{q^{2}+m_{i}^{2}(\phi)/T^{2}})\right).
\label{Vth1}
\end{equation}
The $n_i$ denote the numbers of degrees of freedom present in the 
theory and the upper sign is for bosons and the lower one is for fermions. 
The $\phi$-dependent masses of these degrees
of freedom are denoted as 
 $m_{i}(\phi)$.
In our case there are $n=6$ degrees of freedom corresponding to $W'_{\pm}$ vector bosons of mass
$m(\phi)=g_{\sst \mathrm{D}}\phi$.
In terms of the rescaled dimensionless variables,
\begin{equation}
\gamma \,=\, \phi/w \,, \quad 
\Theta \,=\, T/(g_{\sst \mathrm{D}} w)\,,
\end{equation}
we have,
\begin{eqnarray}
\hat{V}(\gamma,\Theta):= \frac{V_{T}(\phi)}{g_{\sst \mathrm{D}}^4 w^4} =
\frac{3}{32 \pi ^2} \, \gamma^4
\left(-\frac{1}{2}+\log \left(\gamma^2\right)\right) \nonumber\\
+ \frac{6\Theta^4}{2\pi^{2}}\int_{0}^{\infty}\mbox{d}q\, 
q^{2}\log\left(1-\exp(-\sqrt{q^{2}+\gamma^2/\Theta^{2}})\right).
\label{eq:VT1}
\end{eqnarray}
We plot this thermal effective potential in Figs.~\ref{fig:VT1} and  \ref{fig:VT2} as a function of the rescaled scalar field $\gamma \,=\, \phi/w$
for a sequence of temperature values. It easy to see from these figures that there is a barrier separating the two vacua
and thus the phase transition is of the first order. The value
of the critical temperature
where both minima are degenerate and the position of the second minimum are determined numerically to be at\footnote{Note that
unlike in the more familiar SM Higgs effective potential applications, neither the high-temperature nor the low-temperature approximations 
for evaluating $T$-dependence are applicable here.}

\begin{equation}
\Theta_c = \frac{T_c}{g_{\sst \mathrm{D}}w} \simeq 0.312\, , \quad \gamma_c= \frac{\phi_c}{w} \simeq 0.95\,,
\end{equation}
so that the order parameter  $\phi_c/T_c \simeq 3.04/g_{\sst \mathrm{D}} > 1$,  ensuring that
 a {\em first order} phase transition
indeed took place in our weakly coupled model of a dark sector.
This fact is a characteristic feature of Coleman-Weinberg models where the mass parameter at the origin is set to zero
as a consequence of classical scale invariance.

\subsection{Phase transition}
\label{sec:bubbles}

Among the key parameters for the calculation of the gravitational wave spectrum are the rate of variation of the bubble
nucleation rate $\beta$ and the amount of the vacuum energy $\rho_{\rm vac}$ released during the phase transition.
Specifically, following \cite{Grojean:2006bp} we are interested in the dimensionless quantities $\beta/H_*$ 
and $\alpha$ defined below in Eqs.~\eqref{eq:beta} and \eqref{eq:alphadef}.

The thin wall approximation \cite{Coleman:1977py,Anderson:1991zb}
allows for an analytical computation (or estimate) of the parameters characterising the phase transition, and we will consider it
first in Sec.~\ref{sec:thin}. In our model the thin wall approximation, however,
will be seen to break down already at moderately small values of the coupling $g_{\sst \mathrm{D}}\lesssim 1$. 
Therefore we will also consider in Sec.~\ref{sec:triangle} 
a different approximation of the effective potential by a triangular shape.

The probability of bubble formation is proportional to $\exp[-S_4(\phi_{\rm cl})]$ where $S_4$ is the 4-dimensional Euclidean
action corresponding to the tunnelling trajectory and $\phi_{\rm cl}$ is the spherical bubble solution \cite{Kobzarev:1974cp,Coleman:1977py}.
The all-important effects of thermal corrections are taken into account by replacing $S_4$  with the 3-dimensional effective action so that
the probability of tunnelling from a vacuum at the origin $\phi=0$ to the true vacuum $\phi_+$ per unit time per unit volume is 
\begin{equation} 
P = A(T) \exp\left[-S_3(\phi_{\rm cl})/T\right]\, \sim T^4 \exp\left[-S_3(\phi_{\rm cl})/T\right]\,.
\end{equation}
Employing spherical symmetry, the 3D action is 
\begin{equation} 
 S_3 \,=\, 4\pi \int_0^\infty r^2 dr \left(\frac{1}{2}\left(\frac{d\phi}{dr}\right)^2 + V_T(\phi)\right)\,,
\end{equation}
so that the bubble $\phi_{\rm cl}(r)$ configuration is the solution of
\begin{equation}
\frac{d^2 \phi_{\rm cl}}{dr^2} +\frac{2}{r}\frac{d\phi_{\rm cl}}{dr} \,=\, V_T'(\phi_{\rm cl})\,,
\label{eq:bubble}
\end{equation}
with the boundary conditions $\phi_{\rm cl}(\infty)=0$, $d_r\phi_{\rm cl}(0)=0$. 
In the formulae above $V_T$ is the temperature-dependent effective potential \eqref{Vth0}.

After the universe cools down to a temperature below $T_c$ the vacuum at the origin becomes meta-stable, and the bubbles of
true vacuum $\phi_+$ can start appearing. The phase transition occurs when the temperature $T_*$ is reached where the nucleation rate
of the bubbles $P \sim 1$. This occurs when $S_3/T_* \sim 100$.

If this regime can be reached at temperatures just below the critical temperature $T_c$ we would have an $\epsilon$-deviation
from the degenerate vacua. This is depicted by the lowest curve in Fig.~\ref{fig:VT2}. 
Here the parameter $\epsilon$ is the
split in the energy density between the two vacua,
\begin{equation}
\epsilon \,=\, \frac{1}{g_{\sst \mathrm{D}}^4w^4}\, (V_{T}(0) - V_{T}(\phi)) \,.
\label{eq:epsdef}
\end{equation} 
For small $\epsilon$ it is suggestive to employ the thin-wall approximation~\cite{Coleman:1977py,Anderson:1991zb}. To get a first impression of the results this is what we will do in the following. However, we stress here that the smallness of $\epsilon$ is not sufficient for the thin wall approximation to be valid. Indeed the potential barrier as seen from the false vacuum must be large compared to the difference in energy between the true and false vacuum, and this will turn out to be not the case in our model at weak coupling.
Hence we will supplement the thin wall approximation below with a more appropriate treatment in Section~\ref{sec:triangle}.

\subsection{Thin-wall approximation}
\label{sec:thin}

The action in the thin-wall regime is given by the
sum of the volume and the surface terms:
\begin{equation}
S_3\,=\, 4\pi \int_0^R r^2 dr\, V_T(\phi_+)+4\pi R^2\int_0^{\phi_+}\sqrt{2 V_T(\phi)}\, d\phi\,,
\label{eq:thin}
\end{equation} 
where $R$ is the bubble radius and the bubble interpolates between the true vacuum $\phi_+$ for $r<R$ and the 
false $\phi=0$ vacuum at $r>R$. The bubble wall,  $R\pm \delta r$, is thin, $\delta r \ll 1$ for $\epsilon \ll 1$.

The value of the radius $R$ of the bubble is then found by extremising the action $S_3$ with respect to $R$.
For the volume contribution (first term on the {\it r.h.s.} of \eqref{eq:thin}) we have
\begin{equation}
- \, \epsilon g_{\sst \mathrm{D}}^4 w^4 \, \frac{4 \pi}{3} R^3\,,
\end{equation} 
while the surface-tension term gives
\begin{equation}
4 \pi R^2 g_{\sst \mathrm{D}}^2 w^3 \int_0^{\gamma_+}\sqrt{2 V_T(\gamma,\Theta_c)}\, d\gamma
\simeq 4 \pi R^2 g_{\sst \mathrm{D}}^2 w^3 \times 0.0338\,,
\end{equation} 

with the integral having been evaluated numerically.
The bubble radius is found by extremising the action,
\begin{equation}
R\,=\, \frac{2\times 0.0338}{g_{\sst \mathrm{D}}^2 w} \, \frac{1}{\epsilon}\,,
\end{equation} 

and for the action we have,
\begin{equation}
S_3\,=\, \frac{16\pi}{3}\, \frac{(0.0338)^3}{g_{\sst \mathrm{D}}^2 } \, \frac{w}{\epsilon^2}\,.
\end{equation} 

The phase transition completes when
\begin{equation}
\frac{S_3}{T_*}\,\simeq\, \frac{S_3}{T_c}\,=\, \frac{16\pi}{3}\, \frac{1}{0.312}\, \left(\frac{0.0338}{g_{\sst \mathrm{D}}}\right)^3  \, \frac{1}{\epsilon^2} \sim 100\,.
\label{eq:S3est}
\end{equation} 
This implies,
\begin{equation}
\epsilon \simeq \frac{1}{g_{\sst \mathrm{D}}^{3/2}} \, 0.00455\,.
\label{eq:epsest}
\end{equation}

\medskip

We can now compute the $\beta$-parameter characterising the phase transition and in particular the strength of the gravitational wave signal (as we will recall in the next section),
\begin{equation}
\frac{\beta}{H_*}\, =\, T\frac{d}{dT}\left(\frac{S_3}{T}\right)_{T=T_*}.
\label{eq:beta}
\end{equation}
Here $T_*$ is the temperature at which the probability
of nucleating one bubble per horizon volume per unit time is $\sim 1$ (in our case of the thin-wall regime it is
just below $T_c$)
and $H_*$ is the Hubble constant at that time.
A strong gravitational wave signal requires a small $\beta/H_{*}$ so this is the regime we are most interested in.

\begin{figure}[t]
\includegraphics[width=0.43\textwidth]{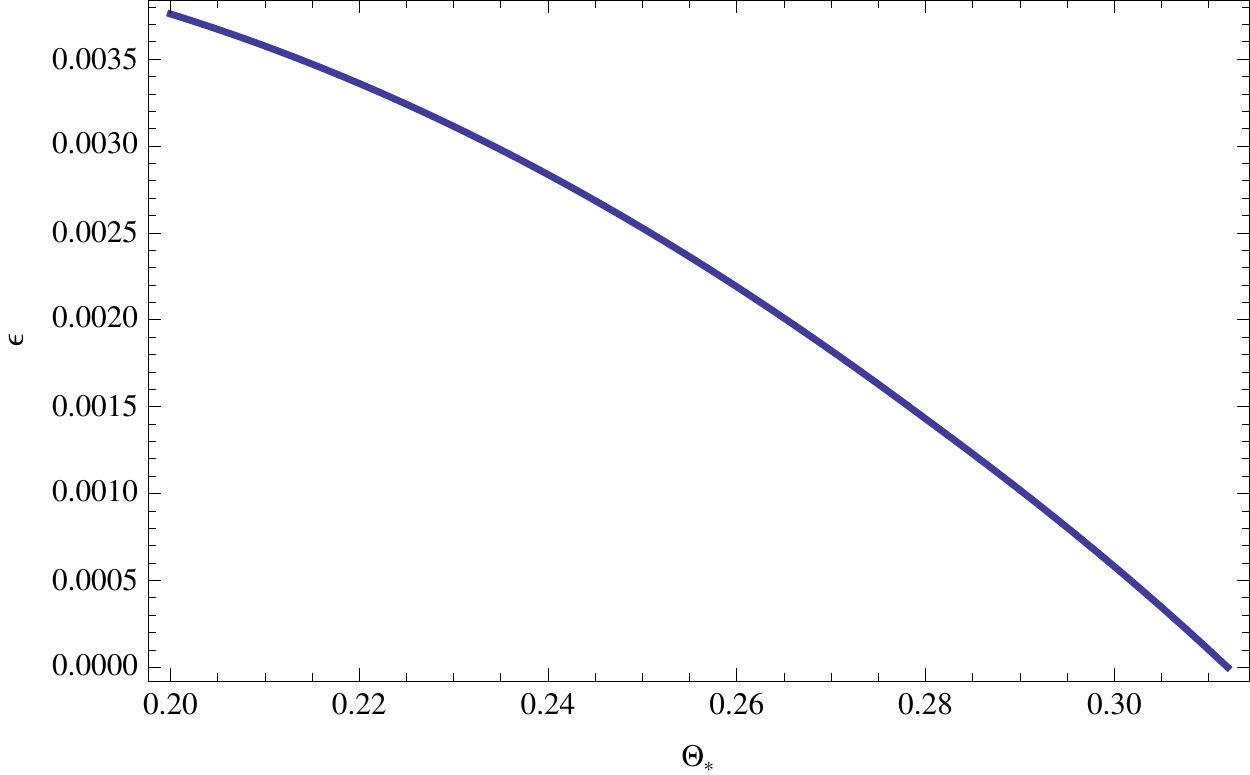}
  \caption{\small $\epsilon$ as a function of the nucleation temperature $T_*$ for $T_* \le T_c$. }
  \label{fig:eps}
\end{figure}

We have computed numerically the dependence of $\epsilon$ on $T$ which is plotted in Fig.~\ref{fig:eps}.
This is very well-described by a numerical fit,
\be
\epsilon (\Theta_*) \simeq\, 
-0.0496 (\Theta_*-0.312) - 
  0.1424 (\Theta_*-0.312)^2
\ee
where $0.312$ is our value for the critical temperature $\Theta_c$.

Now using the expression for the action \eqref{eq:S3est}, the bound $S_3/T_* \simeq 100$ and the fit for $\epsilon (\Theta_*)$ above, 
we find: 
\begin{equation}
\frac{\beta}{H_*}\,=\, \frac{S_3}{T_*}\, \frac{(-2)}{\epsilon} \left(\Theta_* \frac{d \epsilon}{d \Theta_*}\right)_{\Theta_{\rm c}}\, \simeq\,   \frac{3.1}{\epsilon} \,\simeq\,  680\, \, {g_{\sst \mathrm{D}}^{3/2}}\,,
\label{eq:beta}
\end{equation}
where in the final expression we have used Eq.~\eqref{eq:epsest}.

\bigskip 

Finally we need to determine the second key parameter affecting the
gravitational wave spectrum -- the ratio of the vacuum energy density released in the phase transition
to the energy density of the radiation bath,
\begin{equation}
\alpha\,=\, \frac{\rho_{\rm vac}}{\rho_{\rm rad}^*}\,.
\label{eq:alphadef}
\end{equation}
Here $\rho_{\rm rad}^* = g_* \pi^2 T_*^4/30$ and $g_*$ is the number of relativistic degrees of freedom in the plasma at $T_*$.

The vacuum energy, on the other hand, is easy to estimate again in the thin wall approximation as
\begin{equation}
\rho_{\rm vac} \,=\, g_{\sst \mathrm{D}}^4 w^4 \, \epsilon \,\simeq\, 0.00455 \, g_{\sst \mathrm{D}}^{5/2} w^4\,.
\end{equation}
Then we have
\begin{equation}
\alpha\,=\, \frac{1}{g_*\,g_{\sst \mathrm{D}}^{3/2}}\,  \frac{0.137}{\pi^2}\, \frac{1}{\Theta_*^4}\simeq 
\, \frac{1.46}{g_*\, g_{\sst \mathrm{D}}^{3/2}}
\,,
\label{eq:alphaest}
\end{equation}
where we have used $\Theta_* \simeq \Theta_c \simeq 0.312$.

\medskip

As already mentioned above, to safely apply the thin-wall approximation we need not only $\epsilon\ll 1$ but
also $\delta \ll 1$, where we have defined,
\begin{eqnarray}
 \delta& = &\frac{V_{T}(0)-V_{T}(\phi)}{V_{T}(\phi_{max})-V_{T}(0)}
\\\nonumber
&=&\frac{g_{\sst \mathrm{D}}^4w^4}{V_{T}(\phi_{max})-V_{T}(0)}\epsilon
\\\nonumber
&=&\frac{1}{\hat{V}(\gamma_{max},\Theta)}\epsilon,
\end{eqnarray}
and $\phi_{max}=w\gamma_{max}$ is the maximum of the barrier.

As all terms in the potential are dimensionless and arise from 1-loop we generically expect,
\begin{equation}
\hat{V}(\gamma_{max},\Theta)\sim \frac{1}{16\pi^2}.
\end{equation}
This therefore implies,
\begin{equation}
\delta\sim 16\pi^2\epsilon\sim 16\pi^2\frac{0.00455}{g_{\sst \mathrm{D}}^{3/2}}.
\end{equation}
This becomes of order one for $g_{\sst \mathrm{D}}\simeq 0.8$ and the thin wall approximation is problematic
in the weak-coupling regime $g_{\sst \mathrm{D}}\lesssim 1.$

\medskip
\subsection{Beyond the thin-wall approximation}
\label{sec:triangle}

To understand what happens at smaller values of the coupling 
 we adapt the tunnelling approximation of Ref.~\cite{Duncan:1992ai} to the case of our three dimensional thermal bubbles. 
In~\cite{Duncan:1992ai} the authors approximate the potential by a triangle for which the tunnelling solutions can be found analytically.
We will follow this approach to describe the case of broad and low-height barriers we are interested in.

The triangle potential can be characterised by the slope on the left and right hand side of the peak of the triangle, $\lambda_{p}$ and $\lambda_{m}$, as well as the distance between the false vacuum and the top of the potential, $\Delta\phi_{p}$ and the distance from the top to the true vacuum $\Delta\phi_{m}$.
For convenience, as in~\cite{Duncan:1992ai}, we introduce the abbreviations,
\begin{equation}
c=\frac{\lambda_{p}}{\lambda_{m}},\qquad a=(1+c)^{1/3},\qquad \kappa=\frac{\lambda_{p}}{(\Delta\phi_{p})^3}.
\end{equation}

\begin{figure*}[t]
\includegraphics[width=0.4\textwidth]{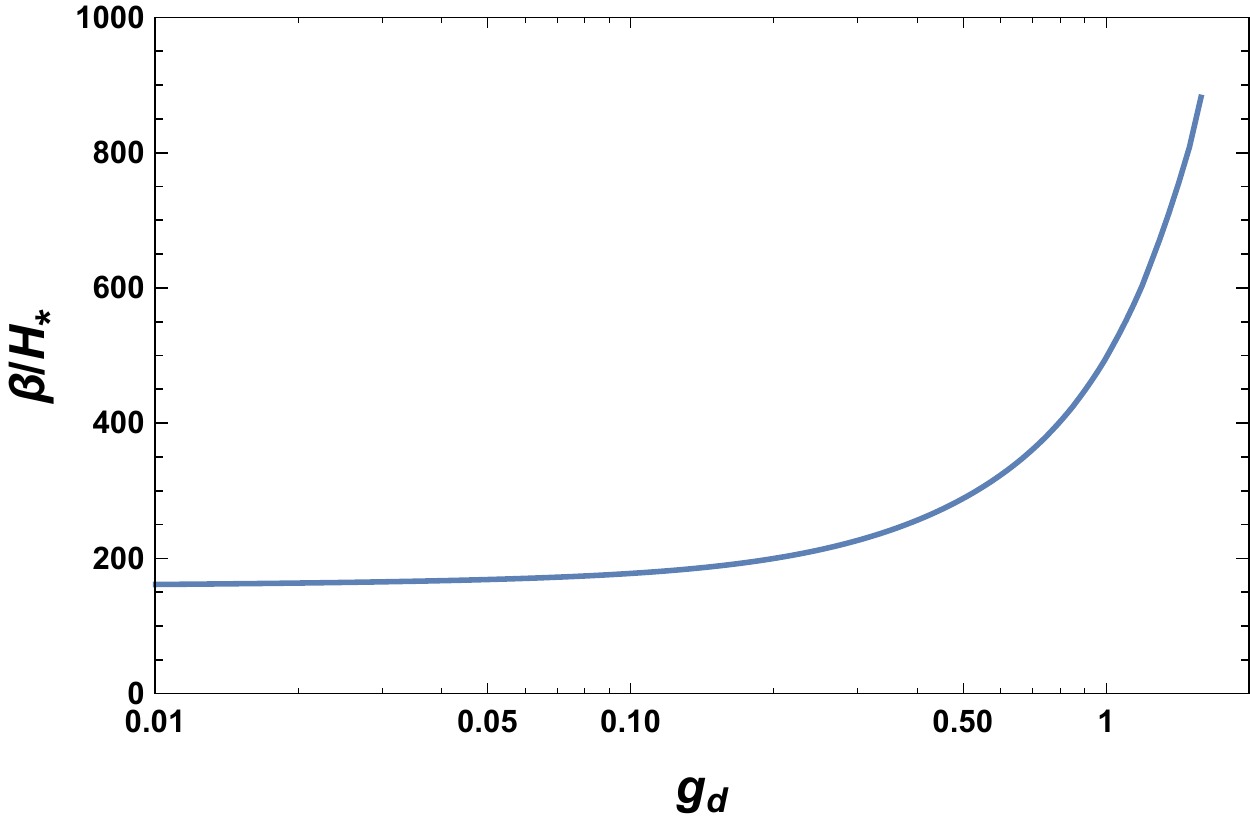}
\hspace{1.0cm}
\includegraphics[width=0.4\textwidth]{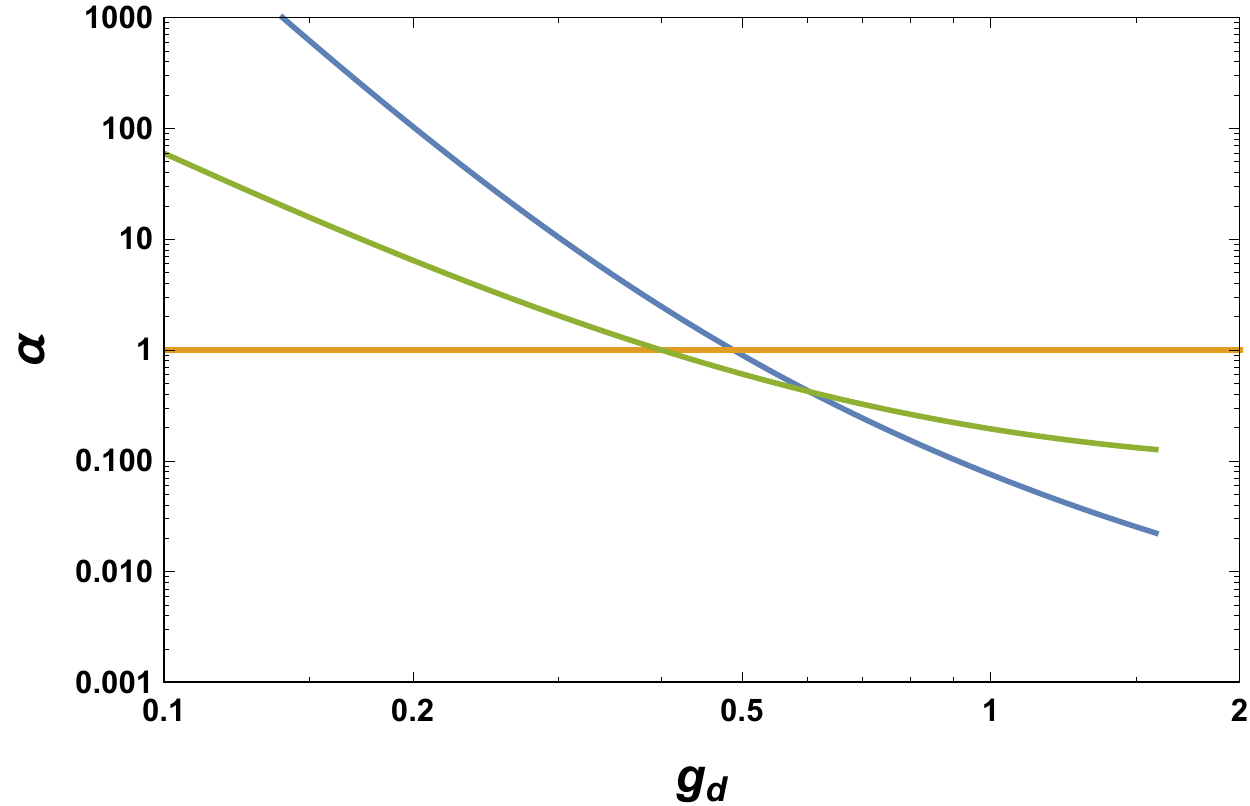}
  \caption{Numerical values for $\beta/H_{*}$ (left) and $\alpha$ (right) for values $g_{\sst \mathrm{D}}\geq 0.1$ in the triangle approximation (blue lines). In the right panel the green line indicates the value of $\alpha_{\infty}$ according to Eq.~\eqref{alphastar} and the golden line indicates $\alpha=1$.}
  \label{betah}
\end{figure*}

The strategy to solve the equation of motion~\eqref{eq:bubble} is as follows. One can easily find solutions to the equations of motion on the right and left hand side of the triangle.
On the right hand side one needs to implement the boundary condition $\phi'(0)=0$. There are two regimes for the field value at $0$. Either the field reaches the true minimum or it does not. The latter happens if $\Delta\phi_{m}$ is sufficiently large. This is what happens for our potential and we will only consider this case in the following. Importantly in this situation there is no dependence on $\Delta\phi_{m}$.
On the left side the field will reach $\phi(R)=0$. Since the potential is linear, $R$ will be finite 
and therefore we also have $\phi'(R)=0$. Finally one can match the two solutions continuously at the top of the triangle. 

After some algebra the result for the 3-dimensional action of the bubble can be written in a relatively compact form as,
\begin{equation}
\label{s3}
S_{3}=\frac{16\sqrt{6} a^3\pi\Delta\phi_{p}}{5\left[(1-a)^2(1+2a)\right]^{2/3}\sqrt{\kappa}}.
\end{equation}

Decreasing the coupling $g_{\sst \mathrm{D}}$, the temperature at which bubbles form also decreases. 
As one can infer from Fig.~\ref{fig:VT1} for smaller temperatures the ratio of the slopes $\lambda_{m}/\lambda_{p}$ goes towards larger values.
It therefore makes sense to approximate Eq.~\eqref{s3} for this case as,
\begin{equation}
\label{s3approx}
S_{3}=\frac{8\sqrt{3}\pi\Delta\phi_{p}}{5\sqrt{c}\sqrt{\kappa}}=\frac{8\sqrt{3}\pi\Delta\phi_{p}^{5/2}}{5\sqrt{\lambda_{m}}}.
\end{equation}

For small temperatures we have checked that to a reasonable approximation the expressions,
\begin{equation}
\Delta\phi_{p}\sim x\Theta w\sim xT/g_{\sst \mathrm{D}},\qquad \lambda_{m}\sim \frac{3}{64\pi^2}g_{\sst \mathrm{D}}^4w^3,
\end{equation}
can be used with 
\begin{equation}
x\sim 0.5-1.2.
\end{equation}
Inserting these formulae into Eq.~\eqref{s3approx} we find,
\begin{equation}
\label{ss3}
\frac{S_{3}}{T}=\frac{64\pi^2}{5g_{\sst \mathrm{D}}^{9/2}} \frac{T^{3/2}}{w^{1/2}}x^{5/2}.
\end{equation}
For the $\beta$ parameter we therefore have,
\begin{equation}
\frac{\beta}{H_{*}}=T\frac{d}{dT}\left(\frac{S_{3}}{T}\right)\bigg|_{T=T_{*}}=\frac{3}{2} \frac{D}{T_{*}}=\frac{3}{2}\frac{S_{3}}{T}\bigg|_{T=T_{*}},
\end{equation}
Since $S_3/T_{*}$ is essentially fixed at$\sim 100$, the same holds for $\beta/H_{*}$ in our model. 
Accordingly we cannot decrease it significantly 
below this value.

\medskip

To complete our estimate we now also need to determine the $\alpha$ parameter in \eqref{eq:alphadef}.
For small temperatures the difference in vacuum energy is simply given by the difference at zero temperature,
\begin{equation}
\rho_{\rm vac}=\frac{3}{64\pi^2}g_{\sst \mathrm{D}}^4w^4.
\end{equation}
Using Eq.~\eqref{ss3} we have for the temperature,
\begin{equation}
T_{*}\sim 0.1 g_{\sst \mathrm{D}}^3\left(\frac{S_{3}}{T_{*}}\right)^{2/3}w.
\end{equation}
This gives,
\begin{equation}
\alpha=\frac{3g_{\sst \mathrm{D}}^4}{64\pi^2}\frac{30}{g_{*}\pi^2 T^{4}_{*}}\sim \frac{60}{g_{*}g_{\sst \mathrm{D}}^8}\left(\frac{S_{3}}{T_{*}}\right)^{-8/3}\sim \frac{0.0003}{g_{*}g_{\sst \mathrm{D}}^8}.
\end{equation}
We stress  that this is a rather crude estimate which is supposed to be valid only for small $g_{\sst \mathrm{D}}\ll 0.1$.

However, there are two messages we can take from this calculation. The first is that with decreasing $g_{\sst \mathrm{D}}$ the transition temperature $T_{*}$ drops dramatically. In line with this the $\alpha$ parameter rapidly increases. 

\bigskip
Finally, 
for larger values of $g_{\sst \mathrm{D}}\geq 0.1$, we have computed the phase transition parameters $\beta/H_{*}$ and $\alpha$ 
numerically, still using the triangle approximation. Their values are plotted in Fig.~\ref{betah}.
We note that for values below $g_{\sst \mathrm{D}}\sim 0.6$, the parameter $\alpha \gtrsim 1$ and the amount of energy in the surrounding plasma is lower than the field energy released in the phase transition. This is important for the gravitational wave signal as we will briefly discuss below.

\subsection{Gravitational waves signal}

\begin{figure*}[t]
\includegraphics[width=0.7\textwidth]{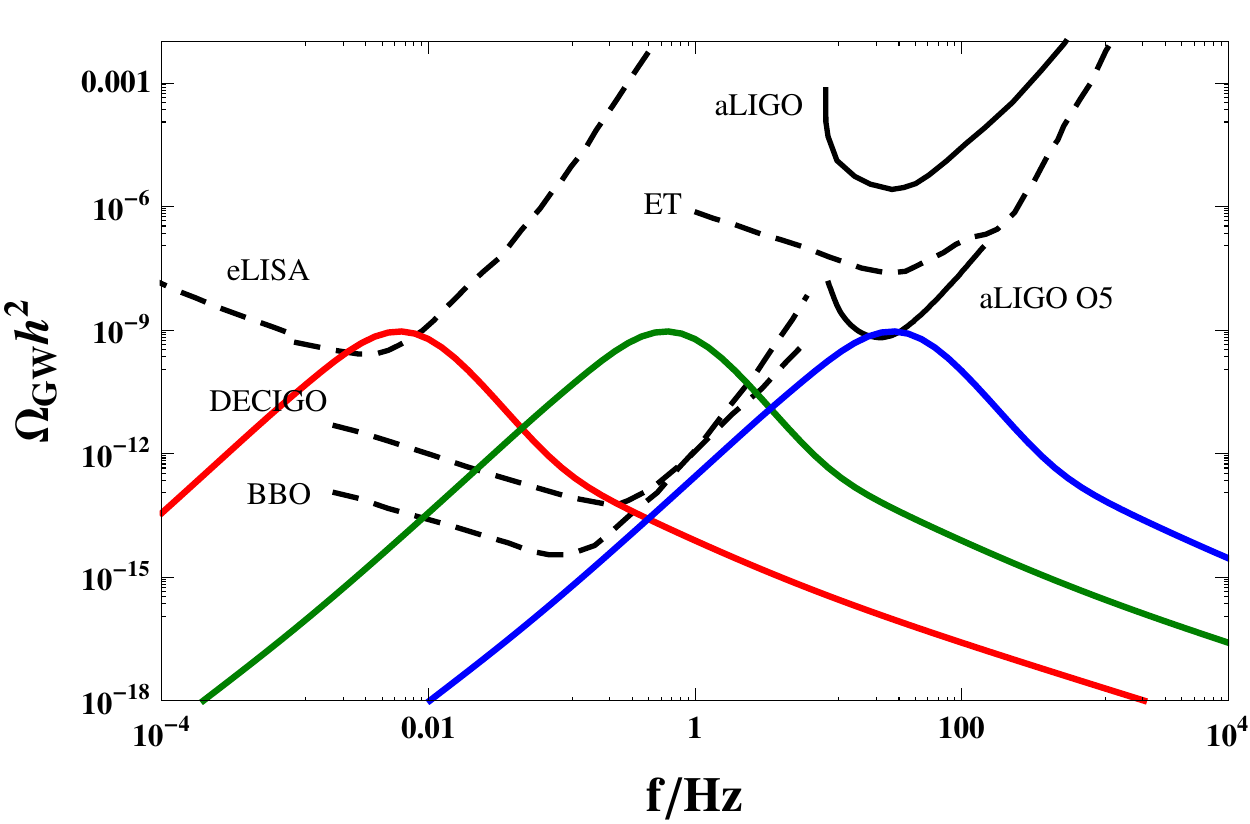}
  \caption{Reach of gravitational wave detectors: We show aLIGO together with the fifth phase of aLIGO (both solid black), and the proposed detectors BBO, DECIGO, ET and eLISA [dashed black] (the sensitivities are taken from the gravitational wave plotter http://rhcole.com/apps/GWplotter/~\cite{Moore:2014lga}). For the curves of the CW phase transition -- going from left to right -- we choose $v_w=1$ throughout, and respectively $(\kappa=1.0, g_D=0.6,T_*=100~\mathrm{GeV})$ [in red], $(\kappa=1.0, g_D=0.6, T_*=10~\mathrm{TeV})$ [green] and $(\kappa=1.0, g_D=0.6, T_*=500~\mathrm{TeV})$ [in blue].}
  \label{gravreach}
\end{figure*}

\begin{figure*}
\includegraphics[width=0.7\textwidth]{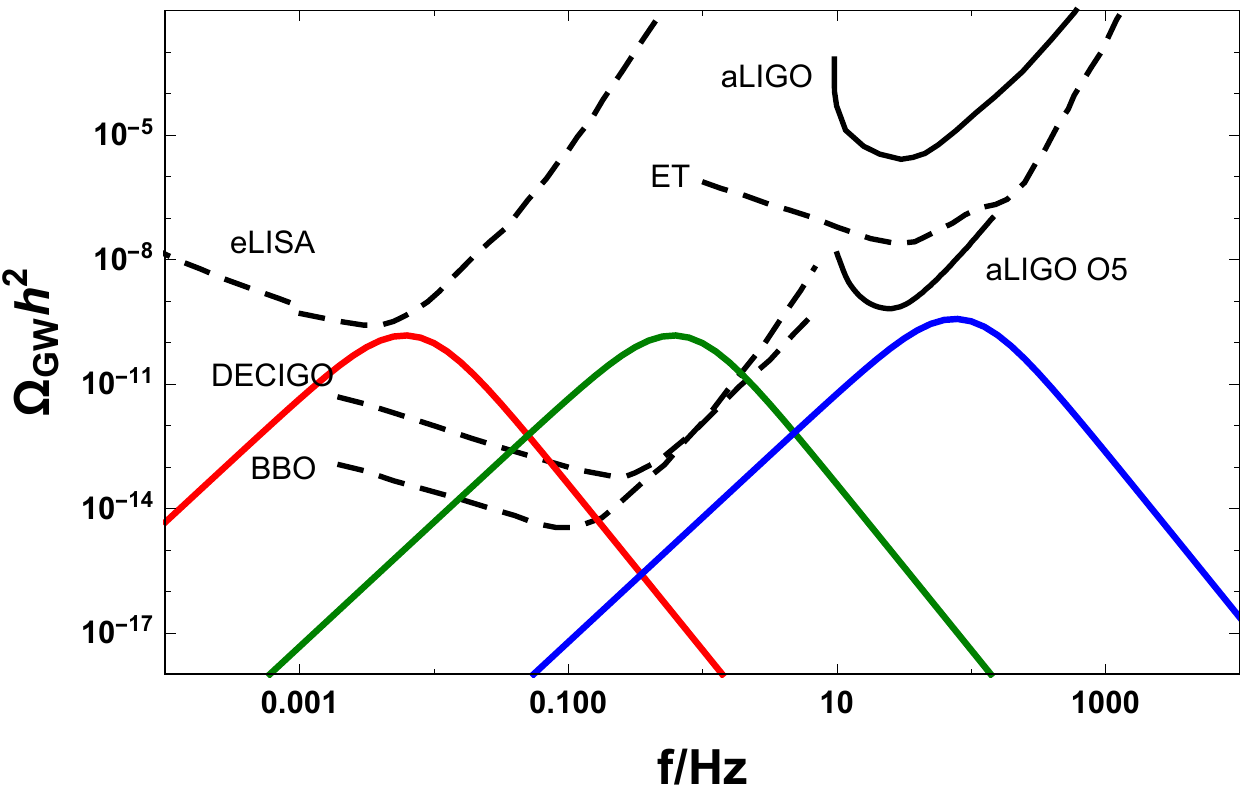}
  \caption{Reach of gravitational wave detectors for a more conservative scenario $\kappa_{sw}=0.4$
  (all other parameters as in Fig.~\ref{gravreach}).}
  \label{gravreach2}
\end{figure*}

As was already discussed and studied in the literature 
\cite{Kosowsky:1991ua,Kamionkowski:1993fg,Grojean:2006bp,Huber:2008hg,Caprini:2009yp,Binetruy:2012ze,Hindmarsh:2015qta,Caprini:2015zlo},  
there are three types of processes during and following the first order phase transition  
involved in the production of gravitational waves: (1) collisions of bubble walls $h^2 \Omega_\mathrm{c}$, (2) sound waves in the plasma $h^2 \Omega_\mathrm{sw}$, and
(3) magnetohydrodynamics turbulence (MHD) following bubble collisions $h^2 \Omega_\mathrm{mhd}$. 

We assume they contribute to the stochastic GW background approximately linearly, i.e.
\begin{equation}
\label{eq:sum}
h^2 \Omega_\mathrm{GW} \simeq h^2 \Omega_\mathrm{c} + h^2 \Omega_\mathrm{sw} + h^2 \Omega_\mathrm{mhd},
\end{equation}
where the three contributions to the signal are given by \cite{Caprini:2015zlo}:
\begin{widetext}
\begin{equation}
 h^2 \Omega_\mathrm{c} = 1.67 \times 10^{-5} \left ( \frac{H_*}{\beta} \right )^2 \left ( \frac{\kappa_\mathrm{c} \alpha}{1+\alpha} \right )^2 \left ( \frac{100}{g_*} \right )^{\frac{1}{3}} \left ( \frac{0.11 v_w^3}{0.42 + v_w^2} \right ) \frac{3.8 (f/f_{\mathrm{env}})^{2.8}}{1+2.8 (f/f_\mathrm{env})^{3.8}},
\end{equation}
\begin{equation}
h^2 \Omega_\mathrm{sw} = 2.65 \times 10^{-6} \left ( \frac{H_*} {\beta} \right ) \left ( \frac{\kappa_\mathrm{sw} \alpha}{1+\alpha} \right )^2 \left ( \frac{100}{g_*} \right )^{\frac{1}{3}} v_w \, \left( \frac{f}{f_{\mathrm{sw}}} \right) \left ( \frac{7}{4+3(f/f_\mathrm{sw})^2} \right )^{7/2} 
\end{equation}
and
\begin{equation}
h^2 \Omega_\mathrm{mhd} = 3.35 \times 10^{-4} \left ( \frac{H_*} {\beta} \right ) \left ( \frac{\kappa_\mathrm{mhd} \alpha}{1+\alpha} \right )^{\frac{3}{2}} \left ( \frac{100}{g_*} \right )^{\frac{1}{3}} v_w\, \frac{(f/f_{\mathrm{mhd}})^3}{\left [ 1 + (f/f_{\mathrm{mhd}})  \right ]^{\frac{11}{3}} (1+8\pi f / h_*)}.
\end{equation}
For the peak frequencies and the Hubble rate after red-shifting for the three processes above we use respectively,
\begin{equation}
f_{\mathrm{env}} = 16.5 \times 10^{-6}~\mathrm{Hz} \left ( \frac{0.62}{1.8-0.1 v_w + v_w^2} \right )  \left ( \frac{\beta}{H_*} \right ) \left ( \frac{T_*}{100~\mathrm{GeV}} \right ) \left ( \frac{g_*}{100} \right )^{\frac{1}{6}},
\end{equation}
\begin{equation}
f_{\mathrm{sw}} = 1.9 \times 10^{-5}~\mathrm{Hz} \left ( \frac{1}{v_w} \right ) \left ( \frac{\beta}{H_*} \right ) \left ( \frac{T_*}{100~\mathrm{GeV}} \right ) \left ( \frac{g_*}{100} \right )^{\frac{1}{6}},
\end{equation}
\begin{equation}
f_{\mathrm{mhd}} = 1.42~f_{\mathrm{sw}}.
\end{equation}
 \end{widetext}
 These expressions depend on the set of key parameters associated with the phase transition: 
 the rate of the phase transition $\beta / H_*$, the energy ratio $\alpha$, 
 together with the latent heat fractions $0<\kappa<1$ for each of the three processes
 and the bubble wall velocity $v_w$. The bubbles are supersonic for $1/\sqrt{3} < v_w \le1$, 
 and subsonic for $v_w \lesssim 1/\sqrt{3}$.
 
 As discussed in Ref.~\cite{Caprini:2015zlo} there are three regimes for the bubbles: non-ruanway bubbles, runaway bubbles in thermal plasma, 
 and runaway bubbles in the vacuum. 
In the non-runaway regime the bubble wall reaches the terminal velocity which remains $v_w<1$.
Such non-runaway bubbles occur for $\alpha<\alpha_{\infty}$, with
\begin{equation}
\label{alphastar}
\alpha_{\infty}\approx\frac{30}{24\pi^2}\frac{\sum_{a} c_{a}\Delta m_{a}^2}{g_{*}T_{*}^2},
\end{equation}
where $c_{a}$ measures the degrees of freedom counting $1$ for bosons and $1/2$ for fermions and $\Delta m_{a}$ is the change 
in the mass of the particles during the phase transition. In this case only the first two mechanisms of gravitational wave production 
contribute, the MHD contribution is absent. For $\alpha\gtrsim \alpha_{\infty}$ it is possible for bubbles to accelerate without bound
(the runaway bubbles) 
and there is no terminal velocity. In this case all three mechanisms contribute into
Eq.~\eqref{eq:sum}. Finally for even larger $\alpha\gg 1$ one is in a situation where the phase transition occurs essentially in vacuum. 
These are runaway bubbles in the vacuum and only the bubble wall collisions processes contribute to the gravitation waves signal.

We find that the signal in general  tends to increase with $\alpha$ and that the sound wave contribution tends to be largest in our model
of the dark sector.
We therefore focus on the case $\alpha\sim \alpha_{\infty}\lesssim 1$\footnote{Here we note some caveats. It is difficult
to pinpoint exactly where the transition between the runaway in the plasma and that in the vacuum occurs. 
Also, the expressions
for $h^2 \Omega$ from~\cite{Caprini:2015zlo} which we use, have only been tested in the $\alpha\lesssim 0.1$ regime~\cite{Caprini:2015zlo}. Our estimates for the signal at $\alpha \sim 0.5$ may therefore be on the optimistic side.}.

For the sound waves the efficiency fraction $\kappa_{sw}$ (for $v_{w}\sim 1$) gives~\cite{Caprini:2015zlo} 
\begin{equation}
\kappa_{sw}\approx\frac{\alpha}{0.73+0.083\sqrt{\alpha}+\alpha}.
\end{equation}
For an example value $\alpha\sim\alpha_{\infty}=0.5$ this is $\sim 0.4$.
Close to the runaway case the colliding bubbles contribution is negligible, and the MHD contribution is typically small, too, $\kappa_{mhd}\sim (0.05-0.1)\kappa_{sw}$ (cf.~\cite{Caprini:2015zlo}).

\bigskip

In Figure~\ref{gravreach} we show the reach of future and current gravitational wave detectors, assuming the optimistic maximal value 
 of $\kappa=1$ for sound waves.
For the number of degrees of freedom we use $g_*=100$. Note, $\Omega_\mathrm{sw} \gg \Omega_\mathrm{c},\Omega_\mathrm{mhd}$ at peak frequency. Over a large part of the parameter space we find good sensitivity at BBO and DECIGO, which cover the frequencies resulting from phase transitions at temperatures of $O(1) \lesssim T_* \lesssim O(10^3)~\mathrm{TeV}$. For even higher frequencies, aLIGO in the fifth phase O5 which is projected to operate in 2020's with design sensitivity taken from Ref.~\cite{TheLIGOScientific:2016wyq}, can also provide sensitivity to phase transitions.

We also show the more conservative case with the lower value of the sound-waves efficiency, $\kappa=0.4$ in Fig.~~\ref{gravreach2}.
Relative to the $\kappa=1$ plots of Fig.~\ref{gravreach}, here we have a loss of sensitivity to aLigo and eLisa experiments.

\section{Domain-wall interactions}
\label{sec:domain}

In models with discrete symmetries domain walls occur quite naturally~\cite{Sikivie:1982qv}. For example they could be formed after a cosmological phase transition where different regions of the Universe settle into different degenerate vacua (connected to each other by the discrete symmetry).

In dark sectors both the distance in field space as well as the height of the potential in between the vacua could be relatively low. In consequence the domain wall tension, i.e. the energy per unit area could be relatively small such that
one could have a reasonable high density of walls without exceeding constraints on the energy density (there have even been suggestion that connect such domain walls to dark matter and dark energy~\cite{Battye:1999eq,Friedland:2002qs}). 

Here we follow the spirit of~\cite{Olive:2010vh,Pospelov:2012mt,Pustelny:2013rza} and consider the observable consequences of the existence of such domain walls. In particular we are interested in signals observable in 
LIGO and other gravitational wave detectors.
While dark sectors by definition are very weakly coupled to Standard Model particles, even low scale domain walls feature relatively large field values. This enhances the signal, making them potentially observable in sensitive experiments.

Interestingly such walls would give distinct transient signals with a variety of shapes (in contrast to the more constant signatures from phase transitions discussed in the previous section.

\subsection*{Domain walls}
Let us consider a domain wall in a pseudo-Goldstone boson which features an additional $Z_{N}$ symmetry.
Following Ref.~\cite{Pospelov:2012mt} we consider the following effective Lagrangian for the domain wall field
\begin{equation}
{\mathcal{L}}_{\phi}=\frac{1}{2}(\partial_{\mu}\phi)^2-2\frac{m^2 f^2}{N^2_{\phi}}\sin^{2}\left(\frac{N_{\phi}\phi}{2f}\right).
\end{equation}
With this the domain wall solutions read,
\begin{equation}
\label{dwsol}
\phi(z)=\frac{4f}{N_{\phi}}\arctan\left[\exp(mz)\right].
\end{equation}

Abundant domain walls would contribute significantly to the energy density. A very conservative constraint is that this contribution should be less than the local dark matter density. Domain walls have a density per unit area $\sigma=mf^2/N^{2}_{\phi}$ and a network with typical distance scale $L$ then has an energy density $\rho\sim \sigma/L$. This gives a limit on the abundance of domain walls~\cite{Pospelov:2012mt},
\begin{equation}
\frac{f}{N_{\phi}}\lesssim  {\rm TeV}\,\times\left(\frac{L}{10^{-2}{\rm Ly}}\right)^{1/2}\left(\frac{{\rm neV}}{m}\right)^{1/2}\left(\frac{\rho_{\rm DW}}{\rho_{DM}}\right)^{1/2}.
\end{equation}
For lower energy densities of the domain wall network one needs a correspondingly lower scale $f$.

Together with the typical velocity $v$ of the domain walls this gives an event rate,
\begin{equation}
{\rm Event\,\,Rate}\sim \frac{1}{10\,\,{\rm years}}\left(\frac{10^{-2}\,{\rm Ly}}{L}\right)\left(\frac{v}{10^{-3}}\right).
\end{equation}
Here the crucial ingredient is the velocity of the domain wall. Inside
the galaxy objects typically have velocities of this order of magnitude and indeed Earth moves with such a velocity around the center of the galaxy. Anything considerably smaller seems a bit fine-tuned.
In principle domain walls could move faster but truly stable ones should be slowed down by the expansion of the Universe\footnote{If the two vacua connected by the domain wall are not exactly equal in energy, the domain wall is in a sense a bubble wall, which could be accelerated by the energy difference and therefore be fast.}. Therefore $v\sim 10^{-3}$ seems a reasonable velocity. 

All in all we want the typical domain wall scale $f$ to be $\lesssim {\rm TeV}$ which is low but still doable.

\subsection*{Interaction with photons}
To have an observable effect in LIGO the domain wall field should have an interaction with Standard Model particles, preferably with photons.
Essentially LIGO measures a phase shift between the two arms of the interferometer.
A simple modification of electrodynamics that leads to a phase shift is a photon mass term inside the domain wall,
\begin{equation}
{\mathcal{L}}_{A}=-\frac{1}{4}F^{\mu\nu}F_{\mu\nu}-\frac{1}{2}m^{2}_{0,\gamma}\sin^{2}\left(\frac{N_{A}\phi}{f}\right)A^{\mu}A_{\mu}.
\end{equation}
Crucially, far away from the plane of the domain wall the effective photon mass is zero in 
agreement with observation, as long as $N_{A}/N_{\phi}$ is integer.

If the photon is effectively massive in some region of space inside the detector this leads to a phase shift.
Approximately one finds\footnote{Here we use a WKB type approximation and neglect reflections on the domain wall. In cavities as employed in LIGO this effect could be non-negligible. Moreover we neglect the small deflection in the propagation direction caused by the domain wall.},
\begin{equation}
\Delta \varphi_{i}=\int_{L_{i}} d{\vec{x}}\, \Delta k({\vec{x}}),
\end{equation}
where $\Delta k({\vec{x}})$ is the space dependent change in wave number and $L_{i}$ denotes the path along the arm $i$ of the interferometer.
The observable quantity is the phase difference between the two paths,
\begin{equation}
\Delta\varphi=\Delta\varphi_{1}-\Delta\varphi_{2}.
\end{equation}

To evaluate this expression we have to determine the change in the wave number in presence of a mass term.
Since the energy of the photon is conserved we have,
\begin{equation}
\Delta k({\vec{x}})=\sqrt{\omega^2-m^{2}_{\gamma}({\vec{x}})}-\omega\approx -\frac{m^{2}_{\gamma}({\vec{x}})}{2\omega},
\end{equation}
where the approximate sign holds for $m_{\gamma}\ll \omega$.
Moreover we we have abbreviated,
\begin{equation}
m^{2}_{\gamma}({\vec{x}})=m^{2}_{0,\gamma}\sin^{2}\left(\frac{N_{A}\phi({\vec{x}})}{f}\right).
\end{equation}
For a completely flat domain wall as in Eq.~\eqref{dwsol} the field value of the wall only depends on the distance to the the wall,
\begin{equation}
\phi({\vec{x}})=\phi({\vec{x}}\cdot{\vec{n}}-z_{0}-vt).
\end{equation}
Here ${\vec{n}}$ is the unit vector normal to the wall, $z_{0}$ is the distance of the wall from the origin at $t=0$ and $v$ is the velocity of the wall with respect to the origin.

\subsection*{Simple examples}
We can choose the arms of the interferometer to be in the $x$ and $y$ direction, respectively.
For simplicity we now take the wall to be parallel to the $z$ direction. Its direction in the $x-y$ plane we specify by
the angle $\alpha$ with respect to the $x$-direction.
For one round trip through the cavity we then obtain the phase shift,
\begin{eqnarray}
\label{shift}
\Delta \varphi(t)\!\!&&\!\!
\\\nonumber
&&\!\!\!\!\!\!\!\!\!\!\!\!\!\!\!\!\!\!\!\!=-\frac{m^{2}_{0,\gamma}}{\omega}\bigg[\int^{L}_{0}\!\!dx\left[\sin^{2}\left(\frac{N_{A}\phi(x\sin(\alpha)-z_{0}-vt)}{f}\right)\right]
\\\nonumber
&&-\int^{L}_{0}\!\!dy\left[\sin^{2}\left(\frac{N_{A}\phi(y\cos(\alpha)-z_{0}-vt)}{f}\right)\right]\bigg].
\\\nonumber
&&\!\!\!\!\!\!\!\!\!\!\!\!\!\!\!\!\!\!\!\!=-\frac{m^{2}_{0,\gamma}}{\omega m}
\\\nonumber
&&\!\!\!\!\!\!\!\!\!\!\!\!\!\!\times
\bigg[\int^{mL}_{0}\!\!d\hat{x}\left[\sin^{2}\left(\frac{N_{A}\phi((\hat{x}\sin(\alpha)-\hat{z}_{0}-v\hat{t})/m)}{f}\right)\right]
\\\nonumber
&&\!\!\!\!\!\!\!\!\!\!\!\!\!\!-\int^{mL}_{0}\!\!d\hat{y}\left[\sin^{2}\left(\frac{N_{A}\phi((\hat{y}\cos(\alpha)-\hat{z}_{0}-v\hat{t})/m))}{f}\right)\right]\bigg],
\end{eqnarray}
where in the second equation we have rescaled to dimensionless variables $\hat{x}=mx,\,\hat{y}=my\,\hat{z}_{0}=mz_{0},\,\hat{t}=mt$. We note that the actual signal is independent of $f$.

The dimensionless mass parameter $m^{2}_{\gamma}/(m\omega)$ controls the overall size of the phase shift. 
The sensitivity of gravitational wave detectors such as LIGO is usually quoted as a sensitivity to a gravitational strain,
\begin{equation}
h_{\rm sens}\sim\frac{\Delta L_{\rm sens}}{L}\sim 10^{-22},
\end{equation}
where $\Delta L_{\rm{sens}}$ is the change in the length of a detector arm caused by the gravitational wave.
In terms of a phase shift for a single path of the detector we therefore have,
\begin{equation}
\label{phisens}
\Delta\varphi_{\rm sens}\sim \Delta L\omega\sim h_{\rm sens}L\omega\sim 10^{-10}.
\end{equation}

\begin{figure}[t]
\includegraphics[width=0.43\textwidth]{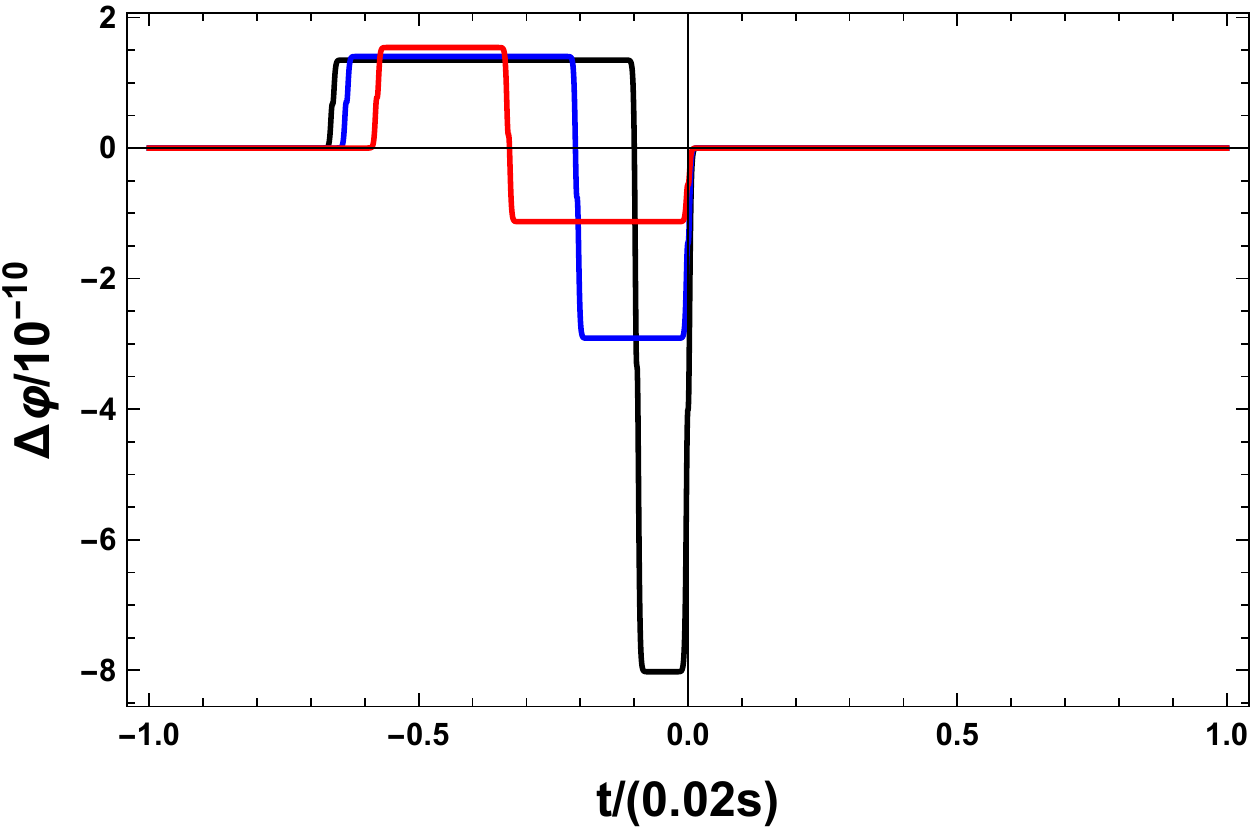}
  \caption{$L=4000\,{\rm m}$, $\omega\approx 1\,{\rm eV}$, $m=10\,{\rm neV}$, $m_{\gamma,0}=1\,{\rm neV}$, $N_{A}/N_{\phi}=1$, $\alpha=\pi/2.2,\pi/2.5,\pi/3$ (black, blue, red), $v$ chosen such that signal has roughly a length of $0.02{\rm s}\sim 1/(50\,{\rm Hz})$ this corresponds to $v= 1\times10^{-3}$.}
  \label{shapes1}
\end{figure}

\begin{figure}[t]
\includegraphics[width=0.43\textwidth]{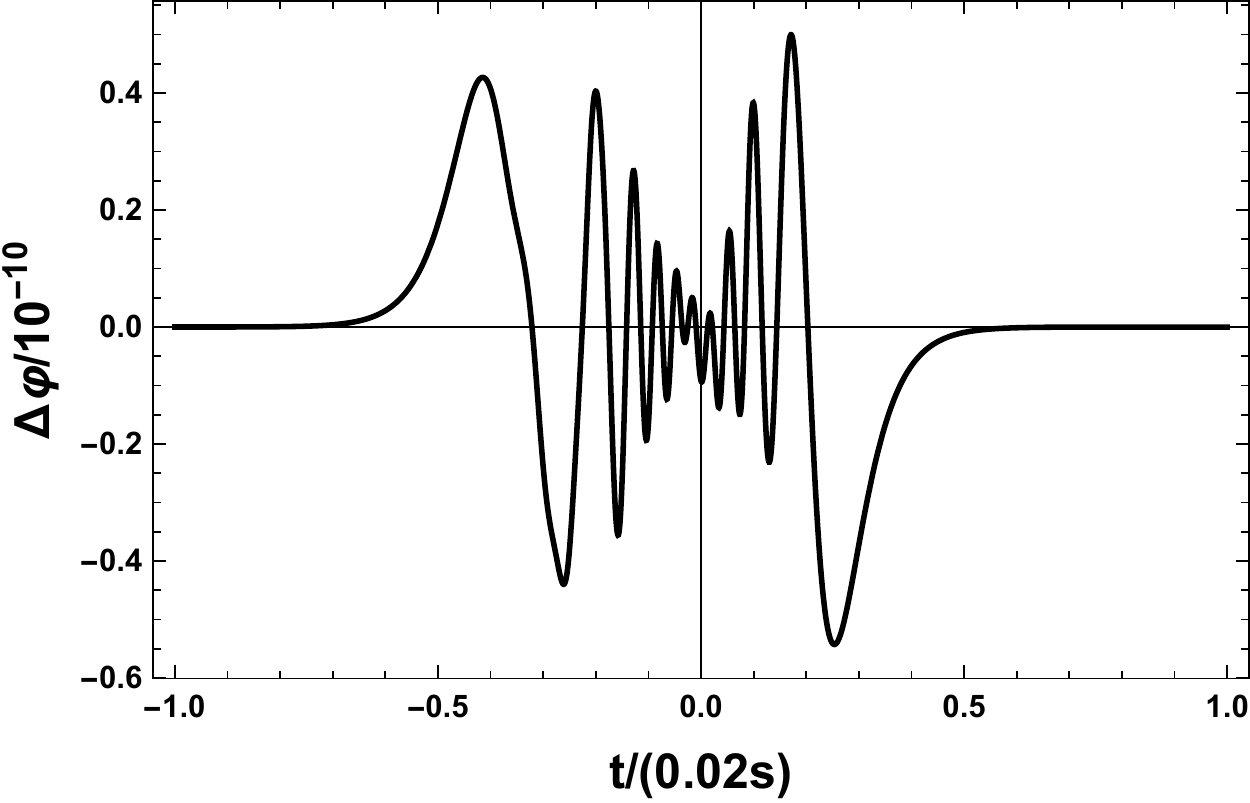}
  \caption{As in Fig.~\ref{shapes1} but $m_{\gamma,0}=0.1\,{\rm neV}$, $N_{A}/N_{\phi}=5$, $m=0.1\,{\rm neV}$, $\alpha=\pi/2.2$ and $v=3\times 10^{-3}$.}
  \label{shapes2}
\end{figure}

\begin{figure}[!t]
\includegraphics[width=0.43\textwidth]{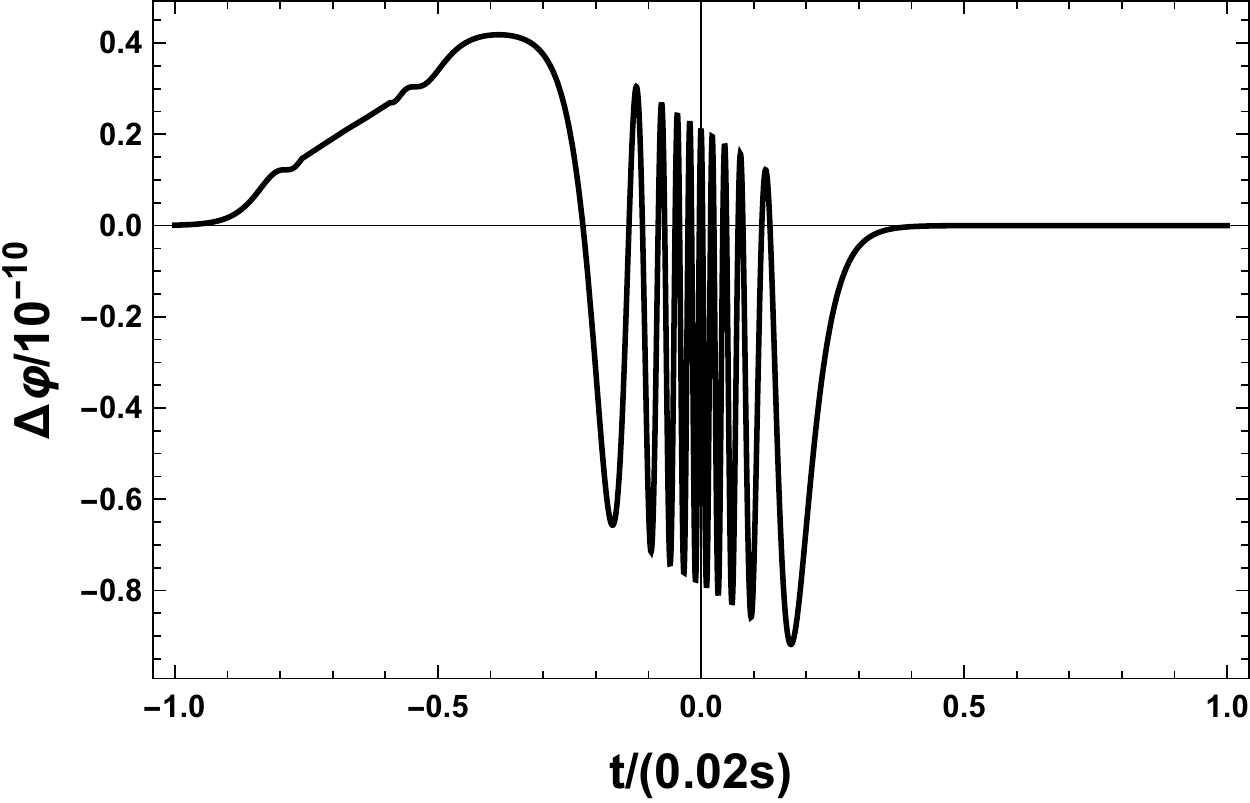}
  \caption{As in Fig.~\ref{shapes1} but $m_{\gamma,0}=0.1\,{\rm neV}$, $N_{A}/N_{\phi}=5$, $m=0.5\,{\rm neV}$, $\alpha=\pi/2$ and $v=1\times 10^{-3}$.}
  \label{shapes3}
\end{figure}

In Figs.~\ref{shapes1},\ref{shapes2},\ref{shapes3} we now show a few different sample shapes that can be produced from these interactions.


From the dimensionless form of Eq.~\eqref{shift} we can determine the typical size of the signal.
The $\sin$ is maximally of order $1$. The region where the $\sin$ is non-vanishing because we are inside the domain wall has length $1$ in these units as well. This allows one to estimate,
\begin{eqnarray}
\Delta\varphi \!\!&\sim&\!\! \frac{m^{2}_{0,\gamma}}{m\omega}\qquad\qquad\qquad\quad\! {\rm for}\,\, mL\gtrsim 1,
\\\nonumber
\!\!&\sim&\!\!   \frac{m^{2}_{0,\gamma}}{m\omega}mL\sim \frac{m^{2}_{0,\gamma}L}{\omega}\quad {\rm for}\,\, mL\lesssim 1.
\end{eqnarray}
For special geometries, where one arm of the detector is essentially parallel to the wall a small enhancement is possible.

Using this and a sensitivity $\Delta\varphi\sim10^{-10}$ we can test the following parameter regions,
\begin{eqnarray}
m_{0,\gamma}\!\!&\sim&\!\! {\rm neV}\left(\frac{m}{10\,{\rm neV}}\right)^{1/2}\quad{\rm for}\,\,m\gtrsim 0.1\,{\rm neV},
\\\nonumber
\!\!&\sim&\!\!  0.1\,{\rm neV}\qquad\qquad\quad\,\,\,{\rm for}\,\, m\lesssim 0.1\,{\rm neV}.
\end{eqnarray}

\subsection*{Signatures of domain wall crossings}
Above we have already seen that domain walls can produce interesting signals which consist of a transient signal with a few oscillations. What is characteristic of those signals and how are they different from gravitational wave signals produced in black hole or neutron star mergers?

The first relevant feature are the typical time-scales and the typical frequencies. 
The duration of the signal is essentially determined by the time it takes the domain wall to cross the detector.
If the wall is thin compared to the size of the detector, i.e. $m\gtrsim 0.1\,{\rm neV}$ this is simply determined
by the length scale of the detector and the velocity of the domain wall,
\begin{equation}
t_{\rm duration}\sim 10\,{\rm ms}\left(\frac{10^{-3}}{v}\right),\quad {\rm thin\,\,wall:}\,\,m\gtrsim 0.1\,{\rm neV}.
\end{equation}
corresponding to frequencies of the order $\sim 100\,{\rm Hz}$.
In addition to the overall length of the signal one will have substructure when the wall enters/leaves one of the arms of the interferometer. The time-scale for this is determined by the thickness of the wall and will have time-scales of the order,
\begin{equation}
t_{\rm substructure}\sim 10\,{\rm ms}\left(\frac{0.1\,{\rm neV}}{m}\right)\left(\frac{10^{-3}}{v}\right),
\end{equation}
corresponding to frequencies $\sim 100\,{\rm Hz}(m/(0.1\,{\rm neV}))$.

For thick walls on the other hand the duration is set by the wall thickness,
\begin{eqnarray}
t_{\rm duration}\!\!&\sim&\!\! 10\,{\rm ms}\left(\frac{0.1\,{\rm neV}}{m}\right)\left(\frac{10^{-3}}{v}\right),
\\\nonumber
&&\qquad\qquad\qquad\qquad{\rm thick \,\,wall:}\,\,m\lesssim 0.1\,{\rm neV}.
\end{eqnarray}

As discussed above the velocity is set by the typical velocities in the galaxies.

The second feature is the time difference between the two detectors at LIGO (or between even more detectors in the future). By the same argument as above this is simply given by the time it takes the domain wall to cross this $\sim 3000\,{\rm km}$ distance,
\begin{equation}
t_{\rm two\,\,detectors}\sim 10\,{\rm s}\left(\frac{10^{-3}}{v}\right).
\end{equation}
This is three orders of magnitude larger than the delay between the signals for gravitational waves. To see a ``coincidence'' one therefore needs to analyze in a suitably large time window.

Indeed one can even perform an additional consistency check between the signals in different locations.
This can be seen most easily in the limit when the wall is thin. Ignoring high frequency substructures the signal then has a shape as in Fig.~\ref{shapes1} which is determined by the angle of the wall with respect to the experiment. Therefore one can measure both velocity and direction of the velocity from a single measurement; the signal for the second site can be predicted.
\bigskip 
\subsection*{Obvious constraints on the parameter space}
Although this is a very simplistic model, let us at least discuss some obvious constraints on the parameter space from other experiments/observations.

{\bf Photons radiating $\phi$:} The mass term for the photon also represents a four boson interaction with coupling strength,
\begin{equation}
\lambda_{AA\phi\phi}\sim \frac{m^{2}_{0,\gamma}N^{4}_{A}}{f^2}\sim 10^{-42}\left(\frac{N_{A}}{1}\right)^{4}
\left(\frac{m_{0,\gamma}}{{\rm neV}}\right)^2\left(\frac{{\rm TeV}}{f}\right)^2.
\end{equation}
It seems like this can be safely ignored.

\bigskip
{\bf Total reflection from the domain wall:} We observe radio signals from very distant astronomical sources in all directions with frequencies down to $\omega\sim (2\pi){\rm few}\,{\rm MHz}\sim {\rm neV}$. If $m_{0,\gamma}\gtrsim {\rm few}\,\,{\rm neV}$ a domain wall would totally reflect all such radio waves, i.e. in the direction where it is coming from we would see no such radio waves.

\subsection*{Beyond the simplest model}
Instead of adding a mass term, one could also consider an axion-like-particle-like interaction of the domain wall with $F^{\mu\nu}F_{\mu\nu}$\footnote{Such an interaction was, e.g. considered in~\cite{Olive:2010vh}.} or $\tilde{F}^{\mu\nu}F_{\mu\nu}$. 
Indeed such a model might be easier to motivate theoretically. Yet the calculation of potential signals (in particular when cavities are employed) needs a more careful study which we leave to future work.

\section{Summary}
\label{sec:conclusion}
In this note we investigated two types of signals from dark sectors observable in gravitational wave detectors: gravitational waves from first order phase transitions and dark sector domain walls very weakly interacting with photons. In the former case future experiments are needed, whereas in the latter case already aLIGO could potentially observe a signal.\\

\section*{Acknowledgements}
\noindent We would like to thank Anupam Mazumdar for interesting discussions on domain walls. JJ gratefully acknowledges support by Transregio TR33 ``The Dark UniverseÕ'
and VVK is supported by the Wolfson foundation. MS and VVK are supported by STFC through the IPPP grant.

%
\newpage
\bibliography{references}

\begin{thebibliography}{35}
\expandafter\ifx\csname natexlab\endcsname\relax\def\natexlab#1{#1}\fi
\expandafter\ifx\csname bibnamefont\endcsname\relax
  \def\bibnamefont#1{#1}\fi
\expandafter\ifx\csname bibfnamefont\endcsname\relax
  \def\bibfnamefont#1{#1}\fi
\expandafter\ifx\csname citenamefont\endcsname\relax
  \def\citenamefont#1{#1}\fi
\expandafter\ifx\csname url\endcsname\relax
  \def\url#1{\texttt{#1}}\fi
\expandafter\ifx\csname urlprefix\endcsname\relax\def\urlprefix{URL }\fi
\providecommand{\bibinfo}[2]{#2}
\providecommand{\eprint}[2][]{\url{#2}}

\bibitem[{\citenamefont{Abbott et~al.}(2016{\natexlab{a}})}]{aLIGO}
\bibinfo{author}{\bibfnamefont{B.~P.} \bibnamefont{Abbott}}
  \bibnamefont{et~al.} (\bibinfo{collaboration}{Virgo, LIGO Scientific}),
  \bibinfo{journal}{Phys. Rev. Lett.} \textbf{\bibinfo{volume}{116}},
  \bibinfo{pages}{061102} (\bibinfo{year}{2016}{\natexlab{a}}),
  \eprint{1602.03837}.

\bibitem[{\citenamefont{Abramovici et~al.}(1992)}]{Abramovici:1992ah}
\bibinfo{author}{\bibfnamefont{A.}~\bibnamefont{Abramovici}}
  \bibnamefont{et~al.}, \bibinfo{journal}{Science}
  \textbf{\bibinfo{volume}{256}}, \bibinfo{pages}{325} (\bibinfo{year}{1992}).

\bibitem[{\citenamefont{Giazotto}(1990)}]{Giazotto:1988gw}
\bibinfo{author}{\bibfnamefont{A.}~\bibnamefont{Giazotto}},
  \bibinfo{journal}{Nucl. Instrum. Meth.} \textbf{\bibinfo{volume}{A289}},
  \bibinfo{pages}{518} (\bibinfo{year}{1990}).

\bibitem[{\citenamefont{Ferdman et~al.}(2010)}]{Ferdman:2010xq}
\bibinfo{author}{\bibfnamefont{R.~D.} \bibnamefont{Ferdman}}
  \bibnamefont{et~al.}, \bibinfo{journal}{Class. Quant. Grav.}
  \textbf{\bibinfo{volume}{27}}, \bibinfo{pages}{084014}
  (\bibinfo{year}{2010}), \eprint{1003.3405}.

\bibitem[{\citenamefont{Harry}(2010)}]{Harry:2010zz}
\bibinfo{author}{\bibfnamefont{G.~M.} \bibnamefont{Harry}}
  (\bibinfo{collaboration}{LIGO Scientific}), \bibinfo{journal}{Class. Quant.
  Grav.} \textbf{\bibinfo{volume}{27}}, \bibinfo{pages}{084006}
  (\bibinfo{year}{2010}).

\bibitem[{\citenamefont{Katz and Riotto}(2016)}]{Katz:2016adq}
\bibinfo{author}{\bibfnamefont{A.}~\bibnamefont{Katz}} \bibnamefont{and}
  \bibinfo{author}{\bibfnamefont{A.}~\bibnamefont{Riotto}}
  (\bibinfo{year}{2016}), \eprint{1608.00583}.

\bibitem[{\citenamefont{Kosowsky et~al.}(1992)\citenamefont{Kosowsky, Turner,
  and Watkins}}]{Kosowsky:1991ua}
\bibinfo{author}{\bibfnamefont{A.}~\bibnamefont{Kosowsky}},
  \bibinfo{author}{\bibfnamefont{M.~S.} \bibnamefont{Turner}},
  \bibnamefont{and} \bibinfo{author}{\bibfnamefont{R.}~\bibnamefont{Watkins}},
  \bibinfo{journal}{Phys. Rev.} \textbf{\bibinfo{volume}{D45}},
  \bibinfo{pages}{4514} (\bibinfo{year}{1992}).

\bibitem[{\citenamefont{Kamionkowski et~al.}(1994)\citenamefont{Kamionkowski,
  Kosowsky, and Turner}}]{Kamionkowski:1993fg}
\bibinfo{author}{\bibfnamefont{M.}~\bibnamefont{Kamionkowski}},
  \bibinfo{author}{\bibfnamefont{A.}~\bibnamefont{Kosowsky}}, \bibnamefont{and}
  \bibinfo{author}{\bibfnamefont{M.~S.} \bibnamefont{Turner}},
  \bibinfo{journal}{Phys. Rev.} \textbf{\bibinfo{volume}{D49}},
  \bibinfo{pages}{2837} (\bibinfo{year}{1994}), \eprint{astro-ph/9310044}.

\bibitem[{\citenamefont{Grojean and Servant}(2007)}]{Grojean:2006bp}
\bibinfo{author}{\bibfnamefont{C.}~\bibnamefont{Grojean}} \bibnamefont{and}
  \bibinfo{author}{\bibfnamefont{G.}~\bibnamefont{Servant}},
  \bibinfo{journal}{Phys. Rev.} \textbf{\bibinfo{volume}{D75}},
  \bibinfo{pages}{043507} (\bibinfo{year}{2007}), \eprint{hep-ph/0607107}.

\bibitem[{\citenamefont{Huber and Konstandin}(2008)}]{Huber:2008hg}
\bibinfo{author}{\bibfnamefont{S.~J.} \bibnamefont{Huber}} \bibnamefont{and}
  \bibinfo{author}{\bibfnamefont{T.}~\bibnamefont{Konstandin}},
  \bibinfo{journal}{JCAP} \textbf{\bibinfo{volume}{0809}}, \bibinfo{pages}{022}
  (\bibinfo{year}{2008}), \eprint{0806.1828}.

\bibitem[{\citenamefont{Caprini et~al.}(2009)\citenamefont{Caprini, Durrer, and
  Servant}}]{Caprini:2009yp}
\bibinfo{author}{\bibfnamefont{C.}~\bibnamefont{Caprini}},
  \bibinfo{author}{\bibfnamefont{R.}~\bibnamefont{Durrer}}, \bibnamefont{and}
  \bibinfo{author}{\bibfnamefont{G.}~\bibnamefont{Servant}},
  \bibinfo{journal}{JCAP} \textbf{\bibinfo{volume}{0912}}, \bibinfo{pages}{024}
  (\bibinfo{year}{2009}), \eprint{0909.0622}.

\bibitem[{\citenamefont{Binetruy et~al.}(2012)\citenamefont{Binetruy, Bohe,
  Caprini, and Dufaux}}]{Binetruy:2012ze}
\bibinfo{author}{\bibfnamefont{P.}~\bibnamefont{Binetruy}},
  \bibinfo{author}{\bibfnamefont{A.}~\bibnamefont{Bohe}},
  \bibinfo{author}{\bibfnamefont{C.}~\bibnamefont{Caprini}}, \bibnamefont{and}
  \bibinfo{author}{\bibfnamefont{J.-F.} \bibnamefont{Dufaux}},
  \bibinfo{journal}{JCAP} \textbf{\bibinfo{volume}{1206}}, \bibinfo{pages}{027}
  (\bibinfo{year}{2012}), \eprint{1201.0983}.

\bibitem[{\citenamefont{Hindmarsh et~al.}(2015)\citenamefont{Hindmarsh, Huber,
  Rummukainen, and Weir}}]{Hindmarsh:2015qta}
\bibinfo{author}{\bibfnamefont{M.}~\bibnamefont{Hindmarsh}},
  \bibinfo{author}{\bibfnamefont{S.~J.} \bibnamefont{Huber}},
  \bibinfo{author}{\bibfnamefont{K.}~\bibnamefont{Rummukainen}},
  \bibnamefont{and} \bibinfo{author}{\bibfnamefont{D.~J.} \bibnamefont{Weir}},
  \bibinfo{journal}{Phys. Rev.} \textbf{\bibinfo{volume}{D92}},
  \bibinfo{pages}{123009} (\bibinfo{year}{2015}), \eprint{1504.03291}.

\bibitem[{\citenamefont{Caprini et~al.}(2015)}]{Caprini:2015zlo}
\bibinfo{author}{\bibfnamefont{C.}~\bibnamefont{Caprini}} \bibnamefont{et~al.}
  (\bibinfo{year}{2015}), \eprint{1512.06239}.

\bibitem[{\citenamefont{Huber et~al.}(2015)\citenamefont{Huber, Konstandin,
  Nardini, and Rues}}]{Huber:2015znp}
\bibinfo{author}{\bibfnamefont{S.~J.} \bibnamefont{Huber}},
  \bibinfo{author}{\bibfnamefont{T.}~\bibnamefont{Konstandin}},
  \bibinfo{author}{\bibfnamefont{G.}~\bibnamefont{Nardini}}, \bibnamefont{and}
  \bibinfo{author}{\bibfnamefont{I.}~\bibnamefont{Rues}}
  (\bibinfo{year}{2015}), \eprint{1512.06357}.

\bibitem[{\citenamefont{Schwaller}(2015)}]{Schwaller:2015tja}
\bibinfo{author}{\bibfnamefont{P.}~\bibnamefont{Schwaller}},
  \bibinfo{journal}{Phys. Rev. Lett.} \textbf{\bibinfo{volume}{115}},
  \bibinfo{pages}{181101} (\bibinfo{year}{2015}), \eprint{1504.07263}.

\bibitem[{\citenamefont{Huang et~al.}(2016)\citenamefont{Huang, Wan, Wang, Cai,
  and Zhang}}]{Huang:2016odd}
\bibinfo{author}{\bibfnamefont{F.~P.} \bibnamefont{Huang}},
  \bibinfo{author}{\bibfnamefont{Y.}~\bibnamefont{Wan}},
  \bibinfo{author}{\bibfnamefont{D.-G.} \bibnamefont{Wang}},
  \bibinfo{author}{\bibfnamefont{Y.-F.} \bibnamefont{Cai}}, \bibnamefont{and}
  \bibinfo{author}{\bibfnamefont{X.}~\bibnamefont{Zhang}}
  (\bibinfo{year}{2016}), \eprint{1601.01640}.

\bibitem[{\citenamefont{Abbott
  et~al.}(2016{\natexlab{b}})}]{TheLIGOScientific:2016wyq}
\bibinfo{author}{\bibfnamefont{B.~P.} \bibnamefont{Abbott}}
  \bibnamefont{et~al.} (\bibinfo{collaboration}{Virgo, LIGO Scientific})
  (\bibinfo{year}{2016}{\natexlab{b}}), \eprint{1602.03847}.

\bibitem[{\citenamefont{Dev and Mazumdar}(2016)}]{Dev:2016feu}
\bibinfo{author}{\bibfnamefont{P.~S.~B.} \bibnamefont{Dev}} \bibnamefont{and}
  \bibinfo{author}{\bibfnamefont{A.}~\bibnamefont{Mazumdar}}
  (\bibinfo{year}{2016}), \eprint{1602.04203}.

\bibitem[{\citenamefont{Coleman and Weinberg}(1973)}]{Coleman:1973jx}
\bibinfo{author}{\bibfnamefont{S.~R.} \bibnamefont{Coleman}} \bibnamefont{and}
  \bibinfo{author}{\bibfnamefont{E.~J.} \bibnamefont{Weinberg}},
  \bibinfo{journal}{Phys. Rev.} \textbf{\bibinfo{volume}{D7}},
  \bibinfo{pages}{1888} (\bibinfo{year}{1973}).

\bibitem[{\citenamefont{Englert et~al.}(2013)\citenamefont{Englert, Jaeckel,
  Khoze, and Spannowsky}}]{Englert:2013gz}
\bibinfo{author}{\bibfnamefont{C.}~\bibnamefont{Englert}},
  \bibinfo{author}{\bibfnamefont{J.}~\bibnamefont{Jaeckel}},
  \bibinfo{author}{\bibfnamefont{V.~V.} \bibnamefont{Khoze}}, \bibnamefont{and}
  \bibinfo{author}{\bibfnamefont{M.}~\bibnamefont{Spannowsky}},
  \bibinfo{journal}{JHEP} \textbf{\bibinfo{volume}{04}}, \bibinfo{pages}{060}
  (\bibinfo{year}{2013}), \eprint{1301.4224}.

\bibitem[{\citenamefont{Khoze et~al.}(2014)\citenamefont{Khoze, McCabe, and
  Ro}}]{Khoze:2014xha}
\bibinfo{author}{\bibfnamefont{V.~V.} \bibnamefont{Khoze}},
  \bibinfo{author}{\bibfnamefont{C.}~\bibnamefont{McCabe}}, \bibnamefont{and}
  \bibinfo{author}{\bibfnamefont{G.}~\bibnamefont{Ro}}, \bibinfo{journal}{JHEP}
  \textbf{\bibinfo{volume}{08}}, \bibinfo{pages}{026} (\bibinfo{year}{2014}),
  \eprint{1403.4953}.

\bibitem[{\citenamefont{Khoze and Ro}(2014)}]{Khoze:2014woa}
\bibinfo{author}{\bibfnamefont{V.~V.} \bibnamefont{Khoze}} \bibnamefont{and}
  \bibinfo{author}{\bibfnamefont{G.}~\bibnamefont{Ro}}, \bibinfo{journal}{JHEP}
  \textbf{\bibinfo{volume}{10}}, \bibinfo{pages}{61} (\bibinfo{year}{2014}),
  \eprint{1406.2291}.

\bibitem[{\citenamefont{Dolan and Jackiw}(1974)}]{Dolan:1973qd}
\bibinfo{author}{\bibfnamefont{L.}~\bibnamefont{Dolan}} \bibnamefont{and}
  \bibinfo{author}{\bibfnamefont{R.}~\bibnamefont{Jackiw}},
  \bibinfo{journal}{Phys. Rev.} \textbf{\bibinfo{volume}{D9}},
  \bibinfo{pages}{3320} (\bibinfo{year}{1974}).

\bibitem[{\citenamefont{Coleman}(1977)}]{Coleman:1977py}
\bibinfo{author}{\bibfnamefont{S.~R.} \bibnamefont{Coleman}},
  \bibinfo{journal}{Phys. Rev.} \textbf{\bibinfo{volume}{D15}},
  \bibinfo{pages}{2929} (\bibinfo{year}{1977}), \bibinfo{note}{[Erratum: Phys.
  Rev.D16,1248(1977)]}.

\bibitem[{\citenamefont{Anderson and Hall}(1992)}]{Anderson:1991zb}
\bibinfo{author}{\bibfnamefont{G.~W.} \bibnamefont{Anderson}} \bibnamefont{and}
  \bibinfo{author}{\bibfnamefont{L.~J.} \bibnamefont{Hall}},
  \bibinfo{journal}{Phys. Rev.} \textbf{\bibinfo{volume}{D45}},
  \bibinfo{pages}{2685} (\bibinfo{year}{1992}).

\bibitem[{\citenamefont{Kobzarev et~al.}(1975)\citenamefont{Kobzarev, Okun, and
  Voloshin}}]{Kobzarev:1974cp}
\bibinfo{author}{\bibfnamefont{I.~{\relax Yu}.} \bibnamefont{Kobzarev}},
  \bibinfo{author}{\bibfnamefont{L.~B.} \bibnamefont{Okun}}, \bibnamefont{and}
  \bibinfo{author}{\bibfnamefont{M.~B.} \bibnamefont{Voloshin}},
  \bibinfo{journal}{Sov. J. Nucl. Phys.} \textbf{\bibinfo{volume}{20}},
  \bibinfo{pages}{644} (\bibinfo{year}{1975}), \bibinfo{note}{[Yad.
  Fiz.20,1229(1974)]}.

\bibitem[{\citenamefont{Duncan and Jensen}(1992)}]{Duncan:1992ai}
\bibinfo{author}{\bibfnamefont{M.~J.} \bibnamefont{Duncan}} \bibnamefont{and}
  \bibinfo{author}{\bibfnamefont{L.~G.} \bibnamefont{Jensen}},
  \bibinfo{journal}{Phys. Lett.} \textbf{\bibinfo{volume}{B291}},
  \bibinfo{pages}{109} (\bibinfo{year}{1992}).

\bibitem[{\citenamefont{Moore et~al.}(2015)\citenamefont{Moore, Cole, and
  Berry}}]{Moore:2014lga}
\bibinfo{author}{\bibfnamefont{C.~J.} \bibnamefont{Moore}},
  \bibinfo{author}{\bibfnamefont{R.~H.} \bibnamefont{Cole}}, \bibnamefont{and}
  \bibinfo{author}{\bibfnamefont{C.~P.~L.} \bibnamefont{Berry}},
  \bibinfo{journal}{Class. Quant. Grav.} \textbf{\bibinfo{volume}{32}},
  \bibinfo{pages}{015014} (\bibinfo{year}{2015}), \eprint{1408.0740}.

\bibitem[{\citenamefont{Sikivie}(1982)}]{Sikivie:1982qv}
\bibinfo{author}{\bibfnamefont{P.}~\bibnamefont{Sikivie}},
  \bibinfo{journal}{Phys. Rev. Lett.} \textbf{\bibinfo{volume}{48}},
  \bibinfo{pages}{1156} (\bibinfo{year}{1982}).

\bibitem[{\citenamefont{Battye et~al.}(1999)\citenamefont{Battye, Bucher, and
  Spergel}}]{Battye:1999eq}
\bibinfo{author}{\bibfnamefont{R.~A.} \bibnamefont{Battye}},
  \bibinfo{author}{\bibfnamefont{M.}~\bibnamefont{Bucher}}, \bibnamefont{and}
  \bibinfo{author}{\bibfnamefont{D.}~\bibnamefont{Spergel}}
  (\bibinfo{year}{1999}), \eprint{astro-ph/9908047}.

\bibitem[{\citenamefont{Friedland et~al.}(2003)\citenamefont{Friedland,
  Murayama, and Perelstein}}]{Friedland:2002qs}
\bibinfo{author}{\bibfnamefont{A.}~\bibnamefont{Friedland}},
  \bibinfo{author}{\bibfnamefont{H.}~\bibnamefont{Murayama}}, \bibnamefont{and}
  \bibinfo{author}{\bibfnamefont{M.}~\bibnamefont{Perelstein}},
  \bibinfo{journal}{Phys. Rev.} \textbf{\bibinfo{volume}{D67}},
  \bibinfo{pages}{043519} (\bibinfo{year}{2003}), \eprint{astro-ph/0205520}.

\bibitem[{\citenamefont{Olive et~al.}(2011)\citenamefont{Olive, Peloso, and
  Uzan}}]{Olive:2010vh}
\bibinfo{author}{\bibfnamefont{K.~A.} \bibnamefont{Olive}},
  \bibinfo{author}{\bibfnamefont{M.}~\bibnamefont{Peloso}}, \bibnamefont{and}
  \bibinfo{author}{\bibfnamefont{J.-P.} \bibnamefont{Uzan}},
  \bibinfo{journal}{Phys. Rev.} \textbf{\bibinfo{volume}{D83}},
  \bibinfo{pages}{043509} (\bibinfo{year}{2011}), \eprint{1011.1504}.

\bibitem[{\citenamefont{Pospelov et~al.}(2013)\citenamefont{Pospelov, Pustelny,
  Ledbetter, Jackson~Kimball, Gawlik, and Budker}}]{Pospelov:2012mt}
\bibinfo{author}{\bibfnamefont{M.}~\bibnamefont{Pospelov}},
  \bibinfo{author}{\bibfnamefont{S.}~\bibnamefont{Pustelny}},
  \bibinfo{author}{\bibfnamefont{M.~P.} \bibnamefont{Ledbetter}},
  \bibinfo{author}{\bibfnamefont{D.~F.} \bibnamefont{Jackson~Kimball}},
  \bibinfo{author}{\bibfnamefont{W.}~\bibnamefont{Gawlik}}, \bibnamefont{and}
  \bibinfo{author}{\bibfnamefont{D.}~\bibnamefont{Budker}},
  \bibinfo{journal}{Phys. Rev. Lett.} \textbf{\bibinfo{volume}{110}},
  \bibinfo{pages}{021803} (\bibinfo{year}{2013}), \eprint{1205.6260}.

\bibitem[{\citenamefont{Pustelny et~al.}(2013)}]{Pustelny:2013rza}
\bibinfo{author}{\bibfnamefont{S.}~\bibnamefont{Pustelny}}
  \bibnamefont{et~al.}, \bibinfo{journal}{Annalen Phys.}
  \textbf{\bibinfo{volume}{525}}, \bibinfo{pages}{659} (\bibinfo{year}{2013}),
  \eprint{1303.5524}.

\end{thebibliography}

\end{document}